\documentclass[superscriptaddress,prd,preprint,notitlepage,nofootinbib,11pt]{revtex4-2}
\usepackage{lmodern}
\usepackage[utf8]{inputenc}
\usepackage[T1]{fontenc}
\usepackage[english]{babel}
\usepackage{graphicx}
\usepackage{caption}
\usepackage{subcaption}
\usepackage[dvipsnames]{xcolor}
\usepackage{tikz}
\usepackage{amsmath,amsfonts,amssymb}
\usepackage{mathtools}
\usepackage{indentfirst}
\usepackage{hyperref}
\usepackage{enumitem}
\usepackage{array}
\usepackage{diagbox}
\usepackage{cases}

\makeatletter
\renewcommand*{\thesection}{\arabic{section}}
\renewcommand*{\thesubsection}{\thesection.\arabic{subsection}}
\renewcommand*{\p@subsection}{}

\renewcommand*{\p@subsubsection}{}
\numberwithin{equation}{section}
\renewcommand*{\theequation}{\thesection.\arabic{equation}}
\renewcommand*{\p@equation}{}
\makeatother

\renewcommand{\Im}{\operatorname{Im}}

\usepackage{hyperref}

\hypersetup{
	colorlinks,%
	citecolor=RoyalPurple,%
	filecolor=blue,%
	linkcolor=ForestGreen,%
	urlcolor=WildStrawberry
}

\begin{document}
\title{Unperturbation theory: \\reconstructing Lagrangians from instanton fluctuations}

\author{Farahmand Hasanov}
\email{khasanov.mkh@phystech.edu}
\affiliation{Moscow Institute of Physics and Technology, 141700, Institutskiy pereulok, 9, Dolgoprudny, Russia}
\affiliation{Theory Department, Lebedev Physics Institute, Leninsky Prospect 53, Moscow 119991, Russia}

\author{Nikita Kolganov}
\email{nikita.kolganov@phystech.edu}
\affiliation{Moscow Institute of Physics and Technology, 141700, Institutskiy pereulok, 9, Dolgoprudny, Russia}
\affiliation{Institute for Theoretical and Mathematical Physics, Moscow State University, 119991, Leninskie Gory, GSP-1, Moscow, Russia}
\affiliation{Theory Department, Lebedev Physics Institute, Leninsky Prospect 53, Moscow 119991, Russia}

\begin{abstract}
	Instantons present a deep insight into non-perturbative effects both in physics and mathematics. While leading instanton effects can be calculated simply as an exponent of the instanton action, the calculation of subleading contributions usually requires the spectrum of fluctuation operator on the instanton background and its Green's function, explicit knowledge of which is rare and a great success. Thus, we propose an inverse problem, namely, the reconstruction of the nonlinear action of the theory admitting instantons from the given fluctuation operator with a known Green's function. We constructively build the solution for this problem and apply it to a wide class of exactly solvable Schr\"{o}dinger operators, called shape-invariant operators, and its simpler subclass, namely reflectionless P\"{o}schl-Teller operators. In the latter case, we found that for the most values of parameters the reconstructed potentials are naturally defined not on the real line, but on some special multisheet covering of the complex plane, and discuss its physical interpretation. For the wider but less simple class of shape-invariant operators, we derive the set of parameters leading to the new infinite families of analytic potentials.
\end{abstract}

\maketitle


\newpage


\section{Introduction}

Instantons are an important concept in various fields of physics from very practical like quantum chromodynamics and solid state physics to almost purely mathematical like matrix models (see \cite{marino2015instantons} for comprehensive review). Their significance is justified by the fact that they provide us with non-perturbative information about the such properties of the system under consideration as partition function, spectrum, correlation functions etc. 

In the most common treatment, instantons are non-trivial saddle points of the Euclidean path integral, representing e.g.\ a partition function of particular quantum field theory model.
Instanton contributions to partition function allow us to find non-perturbative corrections to the energy spectrum, which cannot (at least directly) be obtained from the perturbation theory. Depending on the system under consideration non-perturbative contributions have the meaning of non-perturbative splitting of classically degenerate vacuum energy, decay width of resonant state, width of the energy band in the periodic potential in strong coupling limit, etc \cite{coleman1988aspects,marino2015instantons}. Giving the leading order contribution in the framework of steepest descent method, instantons need to be supplemented by series of corrections, corresponding to the fluctuations about instanton solutions.

Rather unexpected and brilliant observation was that non-perturbative contributions to the energy spectrum are connected to the perturbative ones, e.g.\ the former define the leading degree of divergence of the perturbation theory \cite{bender1973anharmonic,le2012large}. In turn, the corrections, coming from instanton fluctuations, contribute to subleading terms in the behavior of large orders of perturbation theory \cite{Dunne:2015eaa}. The process of calculating these corrections reduces to the calculation of the loop Feynman integrals and involves so-called fluctuation operator, defining the quadratic part of the series expansion of the Euclidean action about instanton solution, Green's function of fluctuation operator, and its functional determinant \cite{Lowe:1978ug,Kolganov:2022ptx}. Except for some specific models, these quantities are not known explicitly, that makes the calculation of the perturbative corrections very complicated.
Thus, one may wonder that there are some alternative ways to access non-perturbative information about the quantum mechanical systems, and extend the aforementioned correspondence of perturbative and non-perturbative physical quantities. 

Indeed, other powerful methods exist, particularly, so-called exact WKB \cite{voros1983return,delabaere1999resurgent,takei2017wkb,Sueishi:2020rug} and uniform WKB methods~ \cite{alvarez2004langer,Dunne:2014bca,Gahramanov:2015yxk}. The former uses the Borel resummation of semiclassical expansion for the wavefunction to recover its global analytic structure, thus allowing to account non-perturbative corrections to the energy spectrum. At the same time, the uniform WKB method exploits the comparison of nearly harmonic behavior of the potential near its minima to the exact solution to the harmonic oscillator Schr\"{o}dinger equation for non-quantized energies in terms of parabolic cylinder functions. Then, one fixes the global boundary conditions to obtain the quantization condition and uses the known asymptotic expansion of the parabolic cylinder to systematically obtain a series for the energy spectrum, containing both perturbative and non-perturbative parts. These methods have a great success for one-dimensional quantum mechanics and are especially important being applied to the Hamiltonians with degenerate minima, for which most other methods fail. Moreover, in some cases it allows to obtain the explicit relations between perturbative and non-perturbative contributions on all orders of perturbation theory.

However, these semiclassical techniques usually deal with the one-dimensional systems, or the systems, that can be reduced to the one-dimensional ones e.g.\ due to a large amount of symmetry. Moreover, WKB methods mostly work in terms of the wavefunctions, thus it seems useless for generalizations to the quantum field theory and calculation of more subtle observables such as many-point correlation functions. Thus, the explicit use of the instanton calculus seems inevitable. 

In order to test the techniques involving instantons, it will be very useful to have a set of toy models, where the calculations can be done as explicit as possible. All the existing examples of the explicit loop calculations rely on the exact knowledge of the Green's function of the fluctuation operator on instanton background. In particular, in one-dimensional quantum mechanical models like double-well or sine-Gordon potential the fluctuation operator itself has the form of the solvable Hamiltonian, namely reflectionless P\"{o}schl-Teller operator, well-known from quantum mechanics textbooks \cite{landau2013quantum,marino2015instantons}. This fact allows to calculate first few corrections to the leading instanton contribution to observable quantities \cite{Lowe:1978ug,Escobar-Ruiz:2015nsa,*Escobar-Ruiz:2015rfa}. However, the calculation of the higher order corrections is still very challenging task.

In this paper we address the following inverse problem: can one reconstruct the full Lagrangian from the given fluctuation operator (defining its quadratic part on the instanton background)? With a positive answer, it becomes possible to obtain (in fact, an infinite) set of the quantum mechanical models that exhibit instanton effects, on the one hand, and has explicitly known Green's function on the other, simply by taking an operator with a known Green's function as a fluctuation operator. We will show that the aforementioned problem indeed can be solved, but the answer has an ambiguity, which can be account in a systematic way.

Our method of the nonlinear potential reconstruction from its fluctuation operator is based on the fact that fluctuation operator on the instanton background always has at least one zero mode, which is the derivative of the instanton solution with respect to the parameter, usually having the meaning of its center. Thus, one can easily reconstruct the instanton solution. Next, using the (Euclidean) energy conservation law, one reconstructs the parametric dependence of the nonlinear potential on the coordinate of the particle. However, this procedure only allows to access the potential in the range of the coordinates which is reached by the instanton trajectory. Thus, one should use some additional information about the system to continue the potential obtained beyond the image of instanton solution, like analyticity or positive definiteness of the potential.

We begin with the observation that most known one-dimensional quantum-mechanical systems, namely double-well, inverted double-well, sine-Gordon and cubic well potentials lead to the fluctuation operators lying in the same class, namely the two-parametric family of reflectionless P\"{o}schl-Teller operators, and differ only by the values of the parameters. The idea is to apply the reconstruction procedure to the reflectionless P\"{o}schl-Teller operators having an arbitrary values of the parameters, in order to obtain an infinite series of the potentials, exhibiting instanton effects. In this direction, we face two problems. The first follows from the observation that if zero mode of the fluctuation operator corresponds the excited state higher then the first one, the reconstructed potential is necessarily a multivalued function, namely it have square root singularities on the instanton trajectory. The second is that for the most values of the parameters, the reconstructed potential turns out to be non-analytic function near the turning points of the instanton trajectory. Thus, we found that for all the parameters of the P\"{o}schl-Teller potentials, excluding ones, corresponding to the fluctuation operators of aforementioned known potentials, our reconstruction technique produces potentials, suffering from nonanalyticity issue. Moreover, we claim that isospectral deformations of these operators do not improve the analytic behavior. Although, we demonstrate that these potentials can be naturally defined on the specific Riemann surfaces of complexified coordinate, we proceed in finding fluctuation operators, leading to analytic reconstructed potentials. For this purpose, we exploit so-called Natanzon operators \cite{Natanzon:1979sr}, which are the most general Schr\"{o}dinger operators, whose solutions can be expressed in terms of hypergeometric function. We apply the reconstruction procedure to Natanzon operators, and then focus on the subclass of so-called shape-invariant operators \cite{Dabrowska:1987fd,Cooper:1986tz,Cooper:1994eh}. Requiring the analyticity of the reconstructed potentials, we fix its parameters, thus finding several infinite families of smooth potentials, whose fluctuation operator on the instanton background are the specific shape invariant operators.

This paper is organized as follows. In Section~\ref{sec:reconstruction} we develop the technique for the reconstruction of the action of the one-dimensional quantum mechanical system, admitting instantons, from given fluctuation operator, defining the quadratic part of the action on the instanton background. Then, in Section~\ref{sec:pt_rec}, we apply the technique obtained to the class of reflectionless P\"{o}schl-Teller operators, thus obtaining two-parametric family of potentials, admitting instantons. We find that these potentials are naturally defined on specific Riemann surfaces, and demonstrate this fact on the instructive examples, corresponding to particular values of the parameters. Also, we discuss the reconstruction of the potentials from the isospectral deformations of the reflectionless P\"{o}schl-Teller operators, and demonstrate that 
In Section~\ref{sec:sip}, we apply the reconstruction technique to a wider class of exactly solvable operators, namely, Natanzon operators, and look for a range of its parameters, leading to the potentials, which are analytic along the real axis. Thus, we find several infinite series of such potentials. Finally, we discuss the results in Section~\ref{sec:discussion}. Appendices~\ref{sec:app_tp_types} and~\ref{sec:app_analytic} contain technical details regarding the analytic behavior of the instanton trajectories, fluctuation operator and the reconstructed potential near the turning points.




\section{Reconstruction procedure} \label{sec:reconstruction}
	Instantons basically arise as nontrivial saddle points of the Euclidean path integral\footnote{Detailed discussion of partition function calculation in the presence of instantons can be found in \cite{marino2015instantons,Kolganov:2022ptx}. Here we mostly focus on the classical part of the story.}
	\begin{equation}
		Z_\beta = \int \mathcal{D} x \; e^{-S[x]}, \label{eq:Eu_PI}
	\end{equation}
	with somehow prescribed boundary conditions in the functional integral measure. Here $S$ is the Euclidean action of the mechanical system, that is considered one-dimensional
	\begin{equation}
		S[x] = \int_{-\beta/2}^{\beta/2} d\tau \, \left(\frac12 \dot x^2 + V(x) \right),
	\end{equation}
	where the dot denotes derivative with respect to Euclidean time $\tau$, whereas $\beta$ is the total Euclidean time of the motion, and $V(x)$ is potential. In turn, instanton solution $x_c(\tau)$ is the solution of the corresponding saddle point equations $\delta S=0$, i.e.\
	\begin{equation}
		\ddot{x}_c(\tau) - V'(x_c(\tau)) = 0, \label{eq:Eu_eom}
	\end{equation}
	having non-trivial dependence on $\tau$, and supplemented with the same boundary conditions as in the path integral measure. Thus, saddle point equation is nothing but Newton's equation for the particle, moving in the inverted potential $-V(x)$.
	
	Depending on the boundary conditions specified, the integral (\ref{eq:Eu_PI}) has the meaning of the partition function, the Euclidean transition amplitude, etc. Here we focus on the case $\beta \to +\infty$, physically corresponding to the zero-temperature partition function, or infinite Euclidean time transition amplitude. Despite the constant solutions $x_c(\tau) = \text{const}$, corresponding to the infinite-time staying at rest in the critical point of the potential $V'(x_c)=0$, there are nontrivial solutions with $\beta\to+\infty$, which can be constructed as follows. Using the Euclidean energy conservation law
	\begin{equation}
		\frac12 \dot{x}^2 - V(x) = E, \label{eq:energy_cons}
	\end{equation} 
	one can integrate e.o.m. (\ref{eq:Eu_eom}), and obtain implicit equation on the trajectory $x(\tau)$ and the classical period $\beta$ as
	\begin{equation}
		\tau = \pm \int^{x(\tau)} \frac{dx}{\sqrt{2(V(x)+E)}}, \qquad \beta = 2 \int_{x_-}^{x_+} \frac{dx}{\sqrt{2(V(x)+E)}},
	\end{equation}
	where $x_\pm$ are the turning points, i.e.\ satisfying $V(x_\pm)+E=0$. The period $\beta$ can be made infinite if the integral diverges due to singularity of the integrand near one or both turning points. Depending on the latter alternative, the corresponding solution has the name of a) tunneling or b) bounce solution (see Fig.~\ref{fig:tun_bnc}). Both are also referred as instantons. In Appendix~\ref{sec:app_tp_types} we demonstrate that for the period $\beta$ to be infinite, the potential should behave as $V(x) \sim |x-x_\pm|^\gamma$,~$\gamma \ge 2$ near one or both turning points, where we set the energy of instanton solution to zero without loss of generality.
	
	\begin{figure}[t]
		\begin{subfigure}[c]{0.4\textwidth}
			\centering
			\includegraphics[scale=1.2]{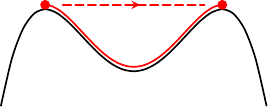}
			\caption{}
		\end{subfigure}	
		\begin{subfigure}[c]{0.4\textwidth}
			\centering
			\includegraphics[scale=1.2]{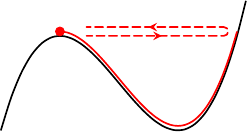}
			\caption{}
		\end{subfigure}
		\caption{Potentials, leading to the two types of typical instanton configurations: (a) tunneling solution, corresponding to infinite-time motion from the one local maximum of the inverted potential $-V(x)$ to another, and (b) bounce solution, consisting of infinite-time motion from the local maximum of the inverted potential, bounce off the turning point, and infinite-time motion back to the local maximum.}
		\label{fig:tun_bnc}
	\end{figure}

	More precisely, \emph{tunneling solution} corresponds to an infinite-time motion from the one local extremum of the inverted potential to another, having the identical values of the potential, so the boundary conditions read
	\begin{equation}
		x_c(+\beta/2) = x_+, \quad x_c(-\beta/2) = x_-, \qquad \beta\to+\infty.
	\end{equation}
	where $x_\pm$ are the turning point, usually coinciding with the local maxima of the inverted potential. In turn, \emph{bounce solution} can be represented as infinite-time motion from the local extremum $x_0$ of the inverted potential, bounce off the turning point $x_b$ at some finite time $\tau_0$, and a subsequent infinite-time motion back to the local maximum. Corresponding boundary conditions are
	\begin{equation}
		x_c(\beta/2) = x_c(-\beta/2) = x_0, \qquad \beta\to+\infty.
	\end{equation}
	\begin{figure}[t]
		\begin{subfigure}[c]{0.49\textwidth}
			\centering
			\includegraphics[scale=1]{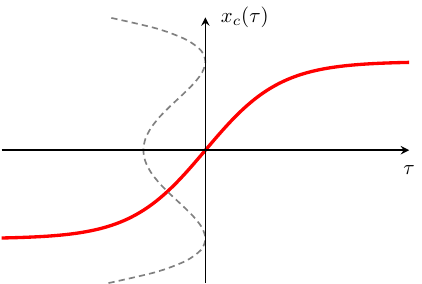}
			\caption{}
		\end{subfigure}	
		\begin{subfigure}[c]{0.49\textwidth}
			\centering
			\includegraphics[scale=1]{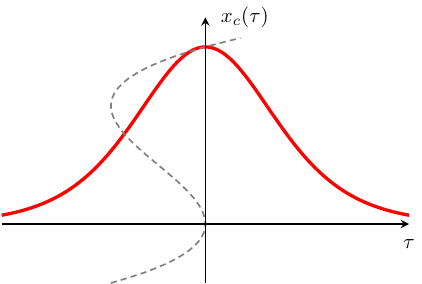}
			\caption{}
		\end{subfigure}
		\caption{Shapes of typical instanton configurations: (a) tunneling solution and (b) bounce solution, corresponding to the infinite Euclidean time motion in the potentials depicted at Fig.~\ref{fig:tun_bnc}.}
		\label{fig:tun_bnc_inst}
	\end{figure}
	See Fig.~\ref{fig:tun_bnc_inst} for typical plots of the tunneling and bounce solutions.
	
\subsection{Fluctuation operators and its zero modes}
	Next, we can expand the action and the corresponding equation of motion about the nontrivial solution $x_c(\tau)$ of the e.o.m.~(\ref{eq:Eu_eom}), e.g.\ bounce or tunneling solution. For this purpose, we define a perturbation $\eta$ about the solution as
	\begin{equation}
		x(\tau) = x_c(\tau) + \eta(\tau),
	\end{equation}
	where $\eta$ obeys the Dirichlet boundary conditions
	\begin{equation}
		\eta(+\beta/2) = \eta(-\beta/2) = 0, \qquad \beta\to\infty. \label{eq:bc_pert_d}
	\end{equation}
	Being expanded about the perturbation $\eta$, the action $S$ takes the following form
	\begin{equation}
		S[x_c+\eta] = S[x_c] + S^{\scriptscriptstyle(2)}[x_c,\eta] + S^{\text{int}}[x_c,\eta],
	\end{equation}
	where $S^{\scriptscriptstyle(2)}$, $S^{\text{int}}$ are called the quadratic action and the interaction action, respectively, and have the following explicit form
	\begin{align}
		S^{\scriptscriptstyle(2)}[x_c,\eta] &= \frac12 \int_{-\beta/2}^{\beta/2} d\tau \; \eta(\tau) \bigl[-\partial_\tau^2 + V''(x_c(\tau))\bigr] \eta(\tau), \\
		S^{\text{int}}[x_c,\eta] &= \sum_{k\ge 3}\frac1{k!}\int_{-\beta/2}^{\beta/2} d\tau \;  V^{(k)}(x_c(\tau)) \, (\eta(\tau))^k.
	\end{align} 
	The quadratic action $S^{\scriptscriptstyle(2)}$ has the following important property. Namely, let us consider the differential operator, defining $S^{\scriptscriptstyle(2)}$
	\begin{equation}
		K = -\partial_\tau^2 + W(\tau), \qquad W(\tau) = V''(x_c(\tau)).
	\end{equation}
	which is called the fluctuation operator and can be identified with the Hamiltonian operator of some one-dimensional quantum mechanical system with a ``coordinate'' $\tau$ and the potential $W(\tau)$. An important fact is that $K$ always has a zero mode $\eta_0(\tau)$, i.e.~$(K\eta_0)(\tau) = 0$. It can be seen by differentiating e.o.m.\ (\ref{eq:Eu_eom}), that immediately gives
	\begin{equation}
		\bigl[-\partial_\tau^2 + V''(x_c(\tau))\bigr] \dot x_c(\tau) = 0
	\end{equation}
	so that $\eta_0(\tau) \propto \dot x_c(\tau)$.
	
	Assuming that the zero-mode is normalizable, we define it as
	\begin{equation}
		\eta_0(\tau) = \frac1{\|\dot x_c\|} \dot x_c(\tau), \qquad \|\dot x_c\|^2 = \int_{-\beta/2}^{\beta/2} dt \, (\dot x_c(\tau))^2. \label{eq:zero_mode_deriv}
	\end{equation}
	In some sense, the fluctuation operator $K$ is in one-to-one correspondence with its zero mode $\eta_0$. Namely, if $K$ is known, then, solving the homogeneous equation
	\begin{equation}
		\bigl[-\partial_\tau^2 + W(\tau)\bigr] \eta_0(\tau) = 0,
	\end{equation}
	with the boundary conditions (\ref{eq:bc_pert_d}), one can obtain $\eta_0$. Conversely, from the known zero-mode, the ``potential term'' of the operator $K$ can be easily found as
	\begin{equation}
		W(\tau) = \frac{\ddot{\eta}_0(\tau)}{\eta_0(\tau)}.
	\end{equation}

	In the purposes of disambiguation, we note that despite the operator $K$, having the form of the Schr\"{o}dinger operator with its own ``potential'' $W(\tau)$, we will not refer to it that way to avoid the confusion of the potential $V(x)$ and $W(\tau)$.

\subsection{Reconstructing potential form fluctuation operator} \label{subsec:rec}
	One can go further and try to reconstruct the whole potential $V(x)$ from the known fluctuation operator, or, equivalently, its zero mode. First of all, using the proportionality (\ref{eq:zero_mode_deriv}) of the zero mode to the corresponding instanton solution, one has
	\begin{equation}
		x_c(\tau) = \nu \int^\tau d\tau' \, \eta_0(\tau'), \label{eq:x_of_tau}
	\end{equation}
	where $\nu = \|\dot x_c\|$ is a normalization factor. Next, one can use the energy conservation law (\ref{eq:energy_cons}) to find the dependence of the potential energy on the Euclidean time on the instanton solution
	\begin{equation}
		V(x_c(\tau)) = \frac12 \dot{x}_c^2(\tau) = \frac{\nu^2}2 (\eta_0(\tau))^2. \label{eq:pot_of_tau}
	\end{equation}
	Thus, one obtains the following parameterization of the potential $V(x)$
	\begin{equation}
		\bigl(x, \, V(x)\bigr) = \left(\nu \int^\tau d\tau' \, \eta_0(\tau'), \; \frac{\nu^2}2 (\eta_0(\tau))^2\right), \qquad -\infty < \tau < + \infty. \label{eq:pot_rec_par}
	\end{equation}
	However this parametric dependence covers only a part of the whole function $V(x)$, namely the image of the instanton solution $x_c(\tau)$. In the case of the tunneling solution it is the interval between two turning points of $V$, i.e.\ $x_- < x < x_+$, whereas for the bounce solution this range is between local extremum $x_0$ of $V$ and the intermediate turning point $x_b$, namely $x_0 < x < x_b$, where $x_b>x_0$ without loss of generality. Thus, one should use some additional information about the system to continue $V(x)$ beyond the image of instanton solution.The possible additional conditions, are e.g.\ analyticity or positive definiteness of the (non-inverted) potential $-V(x)$, which is necessary for the spectrum reality of the underlying quantum mechanical system. Moreover, there is one more ambiguity in the result (\ref{eq:pot_rec_par}), coming from the normalization factor $\nu$, which cannot be found from the fluctuation operator itself. However, this unfixed parameter is not very important, since it can be suppressed by rescaling the coordinate.
	
\subsection{Instructive example: $\delta$-potential}
	\begin{figure}[t]
		\centering
		\begin{subfigure}[c]{0.45\textwidth}
			\centering
			\includegraphics[scale=1]{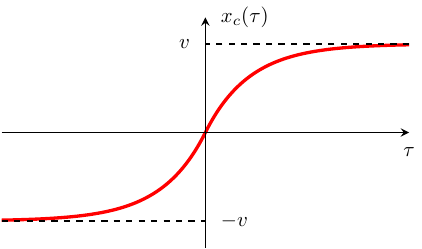}
			\caption{}
			\label{fig:dp_x_of_tau}
		\end{subfigure}	
		\begin{subfigure}[c]{0.45\textwidth}
			\centering
			\includegraphics[scale=1]{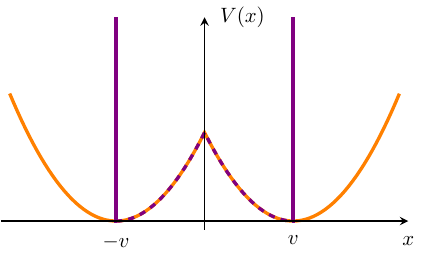}
			\caption{}
			\label{fig:dp_pot_cont}
		\end{subfigure}
		\caption{(a) Shape of the instanton trajectory (\ref{eq:dp_x_of_tau}). (b) Two natural extensions (\ref{eq:dp_pot_cont}) of the potential (\ref{eq:dp_pot_non_cont}) beyond the image of instanton solution: ``analytic'' continuation $V_1(x)$ (orange line), and continuation by infinite walls $V_2(x)$ (violet line).}
	\end{figure}
	Let us apply technique for nonlinear potential reconstructing, described in the previous subsection to a simple example. Namely, let us begin with the following fluctuation operator
	\begin{equation}
		K = -\partial_\tau^2 - 2\varkappa_0 \, \delta(\tau) + \varkappa_0^2, \qquad \varkappa_0 > 0, \label{eq:fl_op_delta_pot}
	\end{equation}
	where $\delta(\tau)$ is usual Dirac delta-function, so that the fluctuation operator $K$ has the form of the Hamiltonian operator of particle with the ``coordinate'' $\tau$ on a line, moving in $\delta$-potential well. The corresponding eigenvalue problem is well-known from quantum mechanics textbooks \cite{griffiths2018introduction}. The only normalizable eigenvector reads
	\begin{equation}
		\eta_0(\tau) = \sqrt{\varkappa_0} \, e^{-\varkappa_0 |\tau|}, \label{eq:dp_zero_mode}
	\end{equation}
	whereas the corresponding eigenvalue is exactly zero due to the constant term $\varkappa_0^{2}$ in (\ref{eq:fl_op_delta_pot}), so that $\eta_0$ is indeed the zero mode. Substitution to (\ref{eq:x_of_tau}) gives the instanton trajectory
	\begin{equation}
		x_c(\tau) = \frac{\nu}{\sqrt{\varkappa_0}} (1-e^{-\varkappa_0 |\tau|}) \operatorname{sign} \tau, \label{eq:dp_x_of_tau}
	\end{equation}
	where we appropriately choose an integration constant (see~Fig.~\ref{fig:dp_x_of_tau}). Thus, the motion of the particle occurs in a finite interval
	\begin{equation}
		-v < x_c(\tau) < v, \qquad 
		v = \frac{\nu}{\sqrt{\varkappa_0}} \label{eq:pt_x_range}
	\end{equation}
	Substituting zero mode to (\ref{eq:pot_of_tau}), and inverting the dependence (\ref{eq:dp_x_of_tau}), one obtains the explicit from of the potential $V(x)$
	\begin{equation}
		V(x) = \frac{\varkappa_0^{2}}2 (v - |x|)^2, \label{eq:dp_pot_non_cont}
	\end{equation}
	for $x$, lying in the range (\ref{eq:pt_x_range}). Thus, one should  artificially continue $V(x)$ beyond this interval somehow. One of the more natural requirements to the continued potential is the positive definiteness of the potential $V(x)$. Thus, there are (at least) two convenient continuation of (\ref{eq:dp_pot_non_cont}) to the whole real line
	\begin{align}
		V_1(x) = \frac{\varkappa_0^{2}}2 (v - |x|)^2, \qquad V_2(x) = \left\{
		\begin{array}{ll}
			\dfrac{\varkappa_0^{2}}2(v - |x|)^2, & |x| < v, \\
			+\infty, & \text{otherwise},
		\end{array}\right. \label{eq:dp_pot_cont}
	\end{align}
	where $V_1(x)$ is just literal (analytic in the vicinity of $x=\pm v$) continuation of the expression (\ref{eq:dp_pot_non_cont}) to the whole real line, whereas $V_2(x)$ is supplementing (\ref{eq:dp_pot_non_cont}) with infinite walls (see~Fig.~\ref{fig:dp_pot_cont}). Thus $V_1(x)$ is nothing but the double oscillator, which is one of the simplest double-well potentials, well-studied in literature \cite{brickmann1975properties,merzbacher1998quantum}.


\section{Reflectionless P\"{o}schl-Teller operators} \label{sec:pt_rec}
	In this section we generalize prototypical examples of the systems with instantons, namely double-well, inverted double-well, sine-Gordon and cubic potentials
	\begin{equation}
		\begin{aligned} \label{eq:pt_anal_pots}
			V_{DW}(x) &=  \frac12(1-x^2)^2, & V_{IDW}(x) &= \frac12 x^2 (1 - x^2), \\ 
			V_{SG}(x) &= 1+\cos(x), & V_{CW}(x) &= 2 \, x^2 (1 - x).
		\end{aligned}
	\end{equation}
	All these potentials admit instanton solutions, having the form
	\begin{equation}
		\begin{tabular}{|c|c|c|c|c|}
			\hline 
			Potential: & Double Well & Inverted DW & Sine-Gordon & Cubic Well \\ \hline 
			$x_c(\tau)$: & $\tanh \tau$ & $\operatorname{sech} \tau$ & $2 \arcsin \tanh\tau$ & $\operatorname{sech}^2\tau $ \\ \hline 
		\end{tabular}
	\end{equation}
	where double-well and sine-Gordon instantons has the form of tunneling solution, whereas cubic-well and inverted double-well instanton is the bounce solution.

	These are united by the fact that the fluctuation operator on the instanton background (up to overall normalization) has the form of P\"{o}schl-Teller operators
	\begin{equation} \label{eq:PT_op}
		K_{\ell,m} = -\partial_\tau^2 + m^2 - \frac{\ell(\ell+1)}{\cosh^2\tau} 
	\end{equation}
	for specific integer values of the parameters:
	\begin{equation} \label{eq:pt_anal_pot_pars}
		\begin{tabular}{|c|c|c|c|c|}
			\hline 
			Potential: & Double Well & Inverted DW & Sine-Gordon & Cubic Well \\ \hline 
			$(\ell, m)$: & $(2, 2)$ & $(2, 1)$ & $(1, 1)$ & $(3, 2)$ \\ \hline 
		\end{tabular}
	\end{equation}
	The P\"{o}schl-Teller operators $K_{\ell,m}$ for an arbitrary integer values of $\ell$,~$m$ are integrable, namely, its eigenvalues, eigenfunctions, and also Green's function are known explicitly \cite{marino2015instantons}. Specifically, for fixed $\ell$, the operator $K_{\ell,m}$ has exactly $\ell$ discrete states, and the corresponding eigenfunctions $\eta_k(\tau)$ and eigenvalues $\lambda_k$ are expressed in terms of the associated Legendre polynomials $P_\ell^n(y)$ as
	\begin{gather}\label{eq:PT_ev}
		\eta_{k}(\tau) = C_{k} P_{\ell}^{m-k}(\tanh\tau), \qquad 
		\lambda_{k} = m^2 - (m-k)^2, \\
		k = m-\ell, \ldots, m-1. \nonumber
	\end{gather}
	Here $C_k$ is the normalization constant (depending also on $\ell$, and $m$), and the range of $k$, numbering the eigenfunctions $\eta_k$, is chosen such that $\eta_0(\tau)$ is exactly zero mode. For the further convenience we provide here the explicit expression for the associated Legendre polynomials
	\begin{equation} \label{eq:legendre_pol}
		P_\ell^n(y) = \frac{(-)^{\ell+n}}{2^\ell \ell!} (1-y^2)^{\frac{n}2} \frac{d^{\ell+n}}{dy^{\ell+n}} (1-y^2)^\ell.
	\end{equation}

	The aim of the present section is to find the potentials $V_{\ell,m}(x)$, whose fluctuation operators on the instanton background are exactly the P\"{o}schl-Teller operators (\ref{eq:PT_op}), for an arbitrary positive integer values of $\ell$,~$m$. To this extent, we apply a simple technique, previously constructed in Section~\ref{subsec:rec}.
	
\subsection{Parametric form of the potentials}
	To reconstruct the potentials $V_{\ell,m}(x)$, from the known fluctuation operator $K_{\ell,m}$, we substitute its zero mode $\eta_0(\tau)$
	\begin{equation} \label{eq:pt_zero_mode}
		\eta_0(\tau) = C_{0} P_{\ell}^{m}(\tanh\tau)
	\end{equation}
	to the expression (\ref{eq:x_of_tau}) for the corresponding instanton solution
	\begin{equation}
		x_c(\tau) = \int^\tau d\tau \, P(\tanh \tau) = \int^{\tanh\tau} \frac{dy}{1-y^2} P(y), \qquad P(y) \coloneqq \nu \,C_0 P_\ell^m(y) \label{eq:PT_inst_sol}
	\end{equation}
	where we change the integration variable from $\tau$ to $y=\tanh \tau$, and introduce the quantity $P(y)$ for the brevity reasons. Thus, the parametric form of the potential $V_{\ell,m}(x)$ reads
	\begin{equation}
		\bigl(x, \, V_{\ell,m}(x)\bigr) = \left(\int^y \frac{dy'}{1-y'^2} P(y'), \; \frac12 P(y)^2\right), \qquad -1 \le y \le 1. \label{eq:PT_pot_par}
	\end{equation}
	As before, this expression gives only the partial information about the potential, namely, its values on the interval, covered by the corresponding instanton solution $x_c(\tau)$. We will show, how to extend $V_{\ell,m}(x)$ beyond this interval in Section~\ref{sec:continuation} below.

	Now, let us examine the expression for the instanton solution, namely the integral on the r.h.s. of (\ref{eq:PT_inst_sol})
	\begin{equation} \label{eq:pt_traj}
		x(y) = \int^y \frac{dy'}{1-y'^2} P(y').
	\end{equation}
	Straightforward analysis shows that $x(y)$ has the following form, depending on the parity of the parameters:
	\begin{equation} \label{eq:x_of_y_parity}
		\begin{tabular}{|c|c|c|c|} \hline
			$\ell$ & $m$ & $x(y)$ & $x(y)$ parity \\ \hline
			Even & Even & $\mathcal{P}_{\ell-1}(y)$ & Odd \\ \hline
			Even & Odd & $\sqrt{1-y^{2}}\,\mathcal{P}_{\ell-2}(y)$ & Even \\ \hline
			Odd & Even & $\mathcal{P}_{\ell-1}(y)$ & Even\\ \hline
			Odd & Odd & $\sqrt{1-y^{2}}\,\mathcal{P}_{\ell-2}(y) + A \arcsin y$ & Odd \\ \hline	
		\end{tabular}
	\end{equation}
	where $\mathcal{P}_n(y)$ is some polynomial of degree $n$ (different for different rows) having definite parity, and $A$ is a constant. For each particular values of $\ell$,~$m$ the function $x(y)$ can be easily recovered. We will discuss the most important examples in Section~\ref{sec:examples}.
	
\begin{figure}[t]
		\begin{subfigure}[c]{0.49\textwidth}
			\centering
			\includegraphics[scale=1]{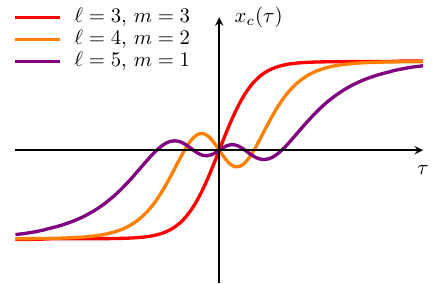}
			\caption{}
		\end{subfigure}	
		\begin{subfigure}[c]{0.49\textwidth}
			\centering
			\includegraphics[scale=1]{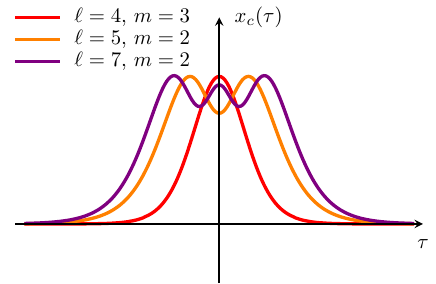}
			\caption{}
		\end{subfigure}
		\caption{Typical shapes of reconstructed instanton trajectories. (i)~even~$\ell-m$, (ii)~odd~$\ell-m$.}
		\label{fig:tun_bnc_inst_PT}
	\end{figure}

	Before proceeding to the continuation of the potentials $V_{\ell,m}(x)$ defined in (\ref{eq:PT_pot_par}) beyond the image of instanton solution, let us focus on the structure of the potential inside this range, and the instanton solutions itself. Depending on its parity, instanton configurations can be classified into bounce-like (odd) and tunneling-like (even) solutions. Thus, from (\ref{eq:x_of_y_parity}) we conclude that even $\ell-m$ correspond to tunneling solutions, whereas odd $\ell-m$ leads to bounce-like solutions. Typical shapes of both classes of the trajectories are depicted at Fig.~\ref{fig:tun_bnc_inst_PT}.
	
\begin{figure}[h!]
		\begin{subfigure}[c]{0.49\textwidth}
			\centering
			\hspace{1em}\includegraphics[scale=1]{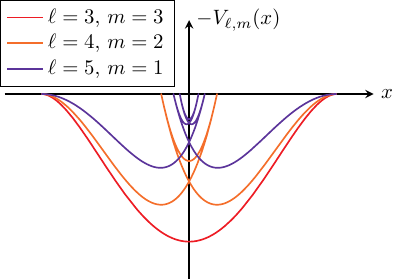}
			\caption{}
		\end{subfigure}	
		\begin{subfigure}[c]{0.49\textwidth}
			\centering
			\includegraphics[scale=1]{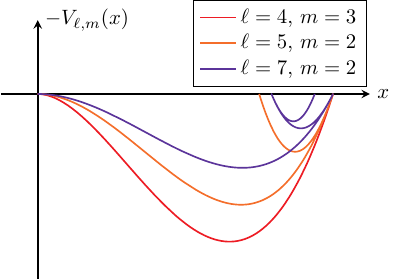}
			\label{fig:bnc_pot_PT}
			\caption{}
		\end{subfigure}
		\caption{Potentials, whose fluctuation operators are reflectionless P\"{o}schl-Teller operators. For $\ell - m > 1$ these are multivalued and have intermediate turning points. (i)~even~$\ell-m$, (ii)~odd~$\ell-m$.}
		\label{fig:tun_bnc_pot_PT}
	\end{figure}

	From the first glance at Fig.~\ref{fig:tun_bnc_inst_PT}, we can see that we are facing a problem. Namely, there are more then one intermediate turning point, that is impossible for convenient classical motion of the particle on the real line. More specifically, there are precisely $\ell-m$ turning points, corresponding to the roots of $P_\ell^m(y)$, excluding the ``boundary'' ones $y=\pm1$. To understand, what is exactly the potentials corresponding to this unusual motion, let us directly apply the formula (\ref{eq:PT_pot_par}) to the case $\ell-m > 1$. It can be easily seen (see Fig.~\ref{fig:tun_bnc_pot_PT}) that the underlying potentials are, in fact, not single-valued functions, but multi-valued ones instead! Specifically, the ``degree of multivaluedness'' for tunneling-like and bounce-like solutions reads
	\begin{equation}
		N_{\text{tunneling}}=\ell-m+1, \qquad
		N_{\text{bounce}} = \frac12(\ell-m+1),
	\end{equation}
	representing the number of the intervals on the $x$ axis glued to each other boundaries. Once an interval is given, the potential on it is uniquely determined. As soon as the particle reaches the boundary of the interval, it moves to another interval with its own potential. However, the total function $V_{\ell,m}(x)$ is still multi-valued, and may (but need not) be single valued only when $m = \ell$ for tunneling case and $m = \ell - 1$ for the bounce one.	This single-valuedness condition will be further refined.

	Let us discuss analytic behavior near the intermediate turning points, i.e.\ finite points of $x$ axis where the derivative of instanton solution vanishes. 
	The tunneling-like and bounce-like solutions have $N_{\text{tunneling}}-1$ and $N_{\text{bounce}}$ such points, respectively.	
	In Appendix~\ref{sec:app_tp_PT} we show that near the turning point at $x = x_{\mathrm{tp}}$ the potential have the following form
	\begin{equation} \label{eq:pt_pot_asympy}
		V_{\ell,m}(x) = (x - x_{\mathrm{tp}}) \sum_{k=0}^{\infty} v_k \, (x - x_{\mathrm{tp}})^{k/2}.
	\end{equation}
	Thus, near general intermediate turning point the potential $V_{\ell,m}(x)$ is linear at the leading order, but the correction terms have branch point singularity at $x=x_{\mathrm{tp}}$. 
	The only exception is the case of odd $\ell-m$ (corresponding to the bounce-like solution), in which exactly one of the turning points is such that the zero mode is an odd function with respect to it. The consequence of this fact is that non-analytic terms disappear from the expression for the potential, i.e.\ $v_{2k+1}=0$ in (\ref{eq:pt_pot_asympy}),
	so it is analytic near this turning point. In fact, this point corresponds to the central point of the bounce-like solution.
	
	Similar to intermediate turning points, let us look at the analytic structure of the potential near boundary turning points, i.e.\ corresponding to the values $y = \pm 1$ in the parameterization (\ref{eq:PT_pot_par}). The analysis is mostly the same as above, despite the fact that, due to the explicit form (\ref{eq:legendre_pol}) of the associated Legendre polynomials, zero mode is not linear but behaves as $(y\mp1)^{2/m}$ near these turning points. As shown in Appendix~\ref{sec:app_tp_PT}, it leads to the following behavior of the potential near the boundary turning points
	\begin{equation} \label{eq:pt_pot_bdy_tp}
		V_{\ell,m}(x) = \frac{m^2}2 (x-x_{\mathrm{tp}})^2 \left(1- \frac{4\ell(\ell+1)}{(m+1)(m+2)}(x-x_{\mathrm{tp}})^{\frac{2}m} + \ldots \right).
	\end{equation}
	where ellipses denote higher powers of $(x-x_{\mathrm{tp}})^{\frac{2}m}$. Thus, for $m>2$ the potential has $2/m$-th root branch point singularity at boundary turning points, and analytic for $m=1,\,2$ near these points. It also worth noting that although the potential is near-harmonic in the vicinity of these turning points, but the anharmonic corrections has the form of series in fractional powers rather then more common Taylor series.
	
	Now, we can formulate the analyticity conditions for the potentials $V_{\ell,m}(x)$. For tunneling solutions, corresponding to even $\ell-m$, intermediate turning points are not allowed, so one should have $N_{\text{tunneling}}=1$, which is satisfied for $\ell=m$. The analyticity near boundary turning points implies $m=1,\,2$, so only cases $(\ell,m)=(1,1),\,(2,2)$ are allowed. In the case of bounce solutions exactly one intermediate turning points is allowed, that implies $N_{\text{bounce}}=1$, so that $\ell-m=1$. Demanding also analytic behavior near the boundary turning points, one obtains that only the values $(\ell,m)=(2,1),\,(3,2)$ are allowed in the bounce-like case. Comparing to (\ref{eq:pt_anal_pot_pars}) we finally conclude that the only \emph{analytic} potentials, having P\"{o}schl-Teller operator as a fluctuation operator on the instanton background, are exactly double-well, inverted double-well, sine-Gordon and cubic-well potentials (\ref{eq:pt_anal_pots}). In what follows, we mostly focus on \emph{non-analytic} cases, and give a physical and mathematical interpretations for them.
	
\subsection{Continuation beyond instanton image} \label{sec:continuation}
	\begin{figure}[h!]
		\centering
		\begin{subfigure}[c]{0.49\textwidth}
			\centering
			\hspace{0.75em}\includegraphics[scale=1]{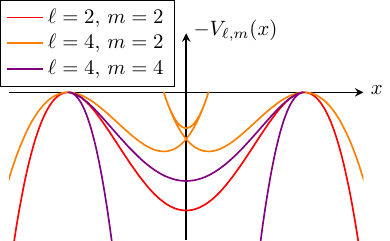}
			\caption{}
		\end{subfigure}	
		\begin{subfigure}[c]{0.49\textwidth}
			\centering
			\includegraphics[scale=1]{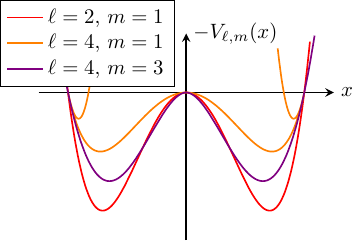}\hspace{0.75em}
			\caption{}
		\end{subfigure}
		\begin{subfigure}[c]{0.49\textwidth}
			\centering
			\includegraphics[scale=1]{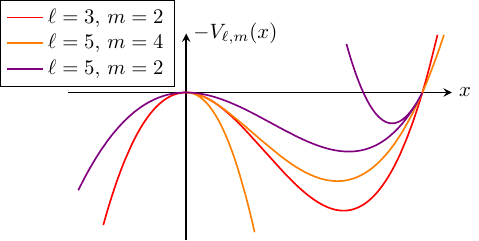}
			\caption{}
		\end{subfigure}
		\begin{subfigure}[c]{0.49\textwidth}
			\centering
			\includegraphics[scale=1]{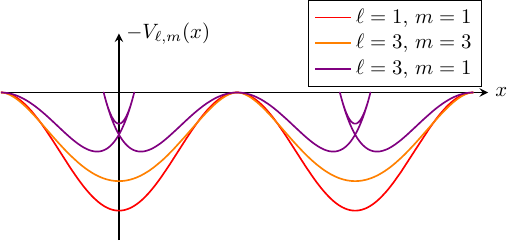}
			\caption{}
		\end{subfigure}
		\caption{
			The potentials, reconstructed from the P\"{o}schl-Teller operators, and analytically continued beyond the range of instanton trajectory. Cases (i)--(iv) correspond to the different parities of the parameters $\ell$ and $m$, summarized in (\ref{eq:pt_parity_table}).
		}
		\label{fig:an_cont}
	\end{figure}
	Now, we proceed to the procedure of continuation beyond the interval, covered by the instanton trajectory. It can be seen from (\ref{eq:x_of_y_parity}) that in the cases, in which $m$ is odd, the parametrically defined potential (\ref{eq:PT_pot_par}) cannot be continued straightforwardly beyond the turning points, corresponding to $y=\pm1$, because of presence of $\sqrt{1-y^2}$ and $\arcsin y$ which are not analytic at these points. Moreover, for odd $\ell$ and even $m$ the instanton trajectory is an even function of the parameter, so straightforward continuation beyond the interval $-1 \le y \le 1$ can only cover the values of the potential on semi-infinite interval of $x$-line.
	Thus, one should provide a more suitable parameterization of instanton trajectory. The main requirement is that both $x$ and $V$ should be analytic functions of the new parameter together with the demanding that $x$ should have at least semi-infinite interval as its image (so, for semi-infinite case one should provide one more parameterization to cover the whole $x$-line).
	
	Let us consider all the cases listed in (\ref{eq:x_of_y_parity}), corresponding to the different parities of $\ell$ and $m$, and provide a suitable parameterization for the continuation beyond the instanton trajectory image.
	\begin{enumerate}[label=(\roman*)]
		\item\label{item:lm_ee} \emph{$\ell$ --- even, $m$ --- even.} This is the simplest case, once a straightforward continuation can be performed. Indeed, in this situation $x(y)$ is an odd analytic polynomial function of $y$, so taking as the range of $y$ not only the interval $-1 \le y \le 1$, but whole real $y$-line, one obtains the values of $V_{\ell,m}(x)$ for all real values of $x$. Moreover, since $(P_\ell^m(y))^2$, defining the potential (cf.\ (\ref{eq:PT_pot_par})) is an even polynomial in $y$, so the potential $V_{\ell,m}(x)$ is bounded from below.
		
		\item\label{item:lm_eo} \emph{$\ell$ --- even, $m$ --- odd.} In this case the function $x(y)$ is not analytic at $y=\pm1$. Nevertheless, after the substitution $y = \sqrt{1-u^2}$, where $u$ is a new parameter, instanton trajectory takes the form $x(y(u)) = u \, \mathcal{Q}_{\ell-2}(u)$, where $\mathcal{Q}_{\ell-2}(u)=\mathcal{P}_{\ell-2}(\sqrt{1-y^{2}})$ is a polynomial in $u$ of degree $\ell-2$ and definite (even) parity. As a consequence, $x(y(u))$ is an odd analytic polynomial function of $u$ and covers the whole $x$-line. The potential $V_{\ell,m}$ is also polynomial as a function of $u$, since $(P_\ell^m(\sqrt{1-u^{2}}))^2$ is a polynomial in $u$ of even degree $2\ell$. Moreover, it can be easily seen that the highest degree term of this polynomial has a negative coefficient, so the potential $V_{\ell,m}(x)$ is bounded from above and unbounded from below. Thus, in this case continuation beyond	boundary turning points can be done with the use of single new parameter $u=\sqrt{1-y^{2}}$.
		\item\label{item:lm_oe} \emph{$\ell$ --- odd, $m$ --- even.} In this situation $x(y)$ is an analytic polynomial function of $y$. However, it is even, so it can cover only semi-infinite interval of real $x$-line. At the same time $u$-parameterization $y=\sqrt{1-u^2}$ give analytic function, namely $x(\sqrt{1-u^2})$ which is also an even polynomial in $u$. To show that these parameterizations cover different pieces of the potential $V_{\ell,m}$, we note that in $y$-parameterization the potential is unbounded from above, whereas in $u$-parameterization $V_{\ell,m}$ is unbounded from below. Indeed, $(P_\ell^m(y))^2$ is even polynomial in $y$ of degree $2\ell$, whose higher degree term has a positive coefficient, whereas $(P_\ell^m(\sqrt{1-u^2}))^2$ is even polynomial in $u$ of degree $2\ell$, but having a negative coefficient of the higher degree term. This situation can take place only if $y$ and $u$ parameterize the different parts of the potential.
		\item\label{item:lm_oo} \emph{$\ell$ --- odd, $m$ --- odd.} For these parities of the parameters, $x(y)$ is non-analytic near $y=\pm1$ due to the presence not only of $\sqrt{1-y^2}$, but also $\arcsin y$ in its explicit form. The parameterization that removes these singularities can be defined as $y=\sin v$. Indeed, the substitution gives $x(y(v)) = \cos v \, \mathcal{P}_{\ell-2}(\sin v) + A \, v$, which is odd analytic function of $v$, so the whole real $x$-line is covered. The potential is also analytic as a function of the new parameter $v$, once $(P_\ell^m(\sin v))^2$ is a polynomial in $\sin v$ and $\cos v$. Immediate consequence of this $v$-parameterization is that the continued potential $V_{\ell,m}(x)$ is periodic.
	\end{enumerate}
	The results of the above analysis can be summarized as the following table
	\begin{equation} \label{eq:pt_parity_table}
			\begin{tabular}{|c|c|c|c|} \hline
				$\ell$ & $m$ & Parameter & Potential \\ \hline
				Even & Even & $y$ & Bounded from below \\ \hline
				Even & Odd & $u = \sqrt{1-y^2}$ & Bounded from above \\ \hline
				Odd & Even & \hspace{1em}$y$,~$u = \sqrt{1-y^2}$ \hspace{1em} & Unbounded \\ \hline
				Odd & Odd & $v = \arcsin y$ & Periodic \\ \hline	
			\end{tabular}
	\end{equation}
	Note that the analytic potentials, namely double-well, inverted double-well, cubic and sine-Gordon potentials correspond exactly to the cases \ref{item:lm_ee}, \ref{item:lm_eo}, \ref{item:lm_oe} and \ref{item:lm_oo}, respectively. Moreover, the potentials inside each class \ref{item:lm_ee}--\ref{item:lm_oo} have a similar shape to those of analytic ``representative'' belonging to each class. For example, the double-well potential, corresponding to the case \ref{item:lm_ee}, in which both $\ell$ and $m$ are even, is bounded from below and unbounded from above, and all other potentials, belonging to this class, exhibit the same property. The representative examples of the continuation procedure application are shown at Fig.~\ref{fig:an_cont}.
	
%
%
%
	
\subsection{Explicit examples} \label{sec:examples}
	In previous subsections we observed that for general parameters $\ell$ and $m$ (excluding those listed in (\ref{eq:pt_anal_pot_pars})) the potential $V_{\ell,m}(x)$ is not analytic functions of $x$, since it has the branch point singularities. Moreover, can be a multi-valued function. Though we examined the behavior of $V_{\ell,m}(x)$ near these singularities, the global analytic structure of the potentials is not described yet. In this subsection we show that allowing $x$ to have complex values, one can treat the potential $V_{\ell,m}(x)$ as a holomorphic function on some Riemann surface, rather than on real line or single complex plane. The motion of the particle in this potential can be viewed as the motion on some (real) section of this surface defined by specific contour (demanding the initial conditions be real). The same contour can be used in the Schr\"{o}dinger equation quantization of the underlying system.
	
	The general analysis is laborious, hence here we limit ourselves to the case where $V_{\ell,m}(x)$ can be found explicitly, i.e.\ the dependence of $x$ in the parameter in (\ref{eq:PT_pot_par}) can be inverted and substituted back to the expression for the potential. This means that the equation $x=x(y)$, where $x(y)$ is defined in (\ref{eq:pt_traj}), can be solved for $y$ in quadratures. This situation takes place, when $x(y)$ has the form of power-law, quadratic, cubic or quartic function of $y$ or a more suitable parameter, e.g. those defined in Section~\ref{sec:continuation}. We will discuss all these cases below.

\subsubsection*{Power-law case: $m = \ell - 1$ }
	\begin{figure}[h!]
		\centering
		\hspace{-2.5em}\begin{subfigure}[c]{0.49\textwidth}
			\centering
			\includegraphics[scale=0.8]{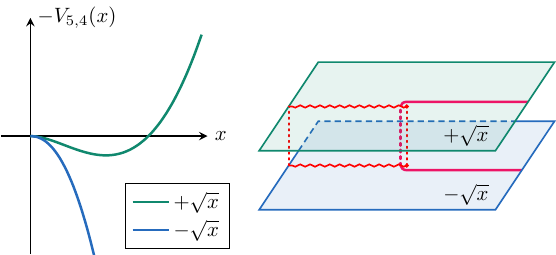}
			\caption{}
		\end{subfigure}
		\begin{subfigure}[c]{0.49\textwidth}
			\centering
			\includegraphics[scale=0.8]{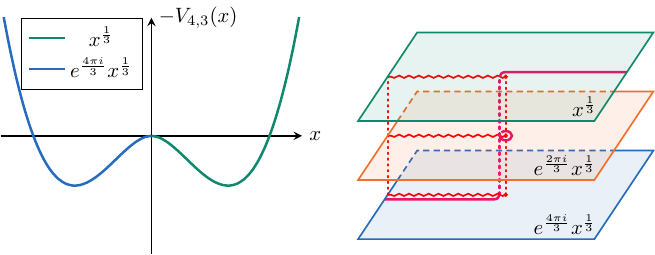}
			\caption{}
		\end{subfigure}
		\caption{Plots of the reconstructed potential $V_{\ell,\ell-1}(x)$ defined in (\ref{eq:pt_pot_l_l1}), and the Riemann surface on which it is analytically continued. Red wiggled lines denote branch cuts, magenta lines denote the contours, corresponding to the real sections. (i) case $\ell=5$, (ii) case $\ell=4$.}
		\label{fig:riemann_exp}
	\end{figure}
	For these values of the parameters, associated Legendre polynomial, defining a zero mode, is simply $P_{\ell}^{\ell-1}(y) = (-)^\ell (2\ell-1)!! \, y \, (1-y^2)^{(\ell-1)/2}$. Thus, the integral (\ref{eq:pt_traj}) can be easily taken, and, after a suitable choice of the normalization coefficient $\nu$, the result reads
	\begin{align}
		P(y) &= - (\ell-1)\, y \left(\frac{1 - y^2}{4}\right)^{\!\!\frac{\ell-1}2} = - (\ell-1)\, \sqrt{1-u^{2}} \, \left(\frac{u}2\right)^{\ell-1}, \\
		x(y) &= \left(\frac{1 - y^2}{4}\right)^{\!\!\frac{\ell-1}2} =\left(\frac{u}2\right)^{\ell-1}, \label{eq:x_of_u_pow_law}
	\end{align}
	where we also go to the parameter $u=\sqrt{1-y^{2}}$, which is convenient, once the parity of parameters under consideration belong to the cases \ref{item:lm_eo} or \ref{item:lm_oe}. Thus, we see that (\ref{eq:x_of_u_pow_law}) is simply power-law equation on the parameter $u$, which can be easily resolved with respect to $u$ and substituted back to $V(y) = P^2(y)/2$ in (\ref{eq:PT_pot_par}). The result reads\footnote{Cf.\ the expansion (\ref{eq:pt_pot_bdy_tp}), which terminates and becomes exact for $m=\ell-1$.}
	\begin{equation} \label{eq:pt_pot_l_l1}
		V_{\ell,\ell-1}(x) = \frac12 (\ell-1)^2\, x^{2}\Bigl(1 - 4 \, x^{\frac{2}{\ell-1}} \Bigr).
	\end{equation}
	Hence, the potential exhibit single ${\frac{2}{\ell-1}}$-th root singularity at $x=0$ for $\ell>3$, so it can be treated as the holomorphic function on the Riemann surface of the function $x^{2/(\ell-1)}$, which is $(\ell-1)$-fold cover of complex plane for even $\ell$ and $(\frac{\ell-1}2)$-fold cover for odd $\ell$, where $x^{2/(\ell-1)}$ has a different phase on each copy. 
	
	In the simplest case of two-fold cover, which take place for $\ell=5$, the potential $V_{5,4}(x) = 8 x^2 \, (1-4\sqrt{x})$ exhibits square root singularity, so the two copies of the complex plane correspond to the two different signs of near the square root. Thus, one can treat this potential as being defined along the contour, going from $+\infty$ to zero on the first copy of the complex plane, passing through the cut to the second one and going to $+\infty$ on the second copy of the complex plane, so that cut crossing correspond to the sing flip of the square root in the expression for the potential (see Fig.~\ref{fig:riemann_exp}).
	
	Similarly, for $\ell=4$ we have the three-fold cover, corresponding to the cube root singularity of the potential $V_{4,3}(x) = 9/2\, x^2 (1-4 x^{2/3}) $. We can define the potential on the contour, going from $+\infty$ to $0$, go around the singularity, and then going to $-\infty$ along the cut (see Fig.~\ref{fig:riemann_exp}). Thus defined potential is real along the contour and symmetric.

\subsubsection*{Biquadratic case: $\ell = 5$, $m=2$}
	\begin{figure}[h!]
		\centering
		\includegraphics[scale=0.8]{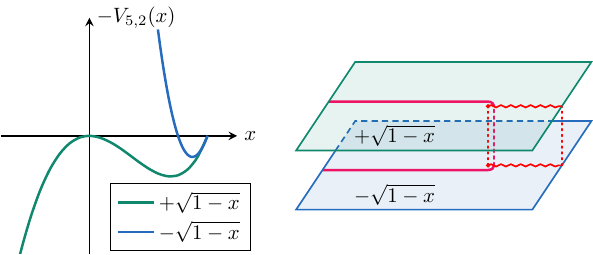}
		\caption{Reconstructed potential $V_{5,2}(x)$ defined in (\ref{eq:pt_pot_52}), and Riemann surface on which it is analytically continued. Red wiggled line denotes branch cuts, magenta line denotes contours, corresponding to real section.}
		\label{fig:riemann_52}
	\end{figure}
	For these values of $\ell$ and $m$ the instanton trajectory is parameterized as
	\begin{align}
		x(y) =  \frac34(1-y^2)(3y^2+1) = -\frac34u^2(3u^2-4), \label{eq:x_of_u_55}
	\end{align}
	so the equation $x = x(y)$ takes the form of biquadratic equation on $y$ or $u$, i.e.\ quadratic equation on $y^2$ or $u^2$, respectively. Zero mode has the form of the polynomial
	\begin{equation}
		P(y) = -3y\,(3y^2-1) = 3\sqrt{1-u^2}\,(3u^2-2),
	\end{equation}
	hence the potential $V_{5,2}(x)$ is an even polynomial on $y^2$ or $u^2$, parameterizing different branches of the potential. Resolving the equation and substituting to the parametric expression for the potential, one obtains
	\begin{equation} \label{eq:pt_pot_52}
		V_{5,2}(x) = -\frac83(1-x)\Bigl[2-3x-2(1-x)^{3/2}\Bigr],
	\end{equation}
	where the different branches of the potential have the different signs of the square root. Thus, the potential has the square root singularity at $x=1$ so it is the holomorphic function on the two-fold cover of the complex plane with the cut, going from $1$ to $\infty$. Thus, the contour, on which the potential is real, and its analytic structure is similar to those of $\ell=5$,~$m=4$ case, described above. An interesting feature of this particular case is that the corresponding instanton trajectory starts on the one copy of the complex plane, then goes to the second one by crossing the cut at first one, reaches the intermediate turning point, and finally returns to infinity by the same trajectory (cf. the plot of appropriate instanton trajectory at Fig.~\ref{fig:tun_bnc_inst_PT}).

\subsubsection*{Cubic case: $\ell=4$, $m = 4,\,2,\,1$}
\begin{figure}[h!]
	\centering
	\includegraphics[scale=0.8]{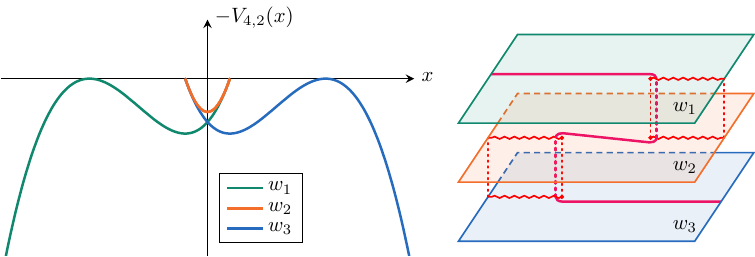}
	\caption{Reconstructed potential $V_{4,2}(x)$ defined in (\ref{eq:pt_pot_42}), and Riemann surface on which it is analytically continued. Red wiggled line denotes branch cuts, magenta line denotes the contours, corresponding to real section.}
	\label{fig:riemann_cubic}
\end{figure}

For these values of $\ell$ and $m$ the equation $x = x(y)$ is cubic equation on $y$ or $u$, namely
\begin{align}
	(4,4):& & x(y) &= \frac12 y \, (y^2-3), \\[0.25em]
	(4,2):& & x(y) &= \frac14y\,(7y^2-3), \\[0em]
	(4,1):& & x(y) &= -\frac{\sqrt{7}}{16}u\,(7u^2-12).
\end{align}
Moreover, these are so-called depressed cubic equations, i.e.\ having no quadratic term. For this case the roots are expressed in terms three roots $w_{1,2,3}$ of the equation
\begin{equation}
	w^3 = x+\sqrt{x^2-1},
\end{equation}
so that the expressions for the potentials read
\begin{align}
	V_{4,4}(x) &= \frac98 \bigl(1 - y^2 \bigr)^4, & y &= -\bigl(w + w^{-1}\bigr),\\
	V_{4,2}(x) &{}= \frac{63}8(7y^4-8y^2+1)^2, & y &= w+w^{-1}, \label{eq:pt_pot_42}\\
	V_{4,1}(x) &= \frac{63}{512} u^2 (1-u^2)(4-7u^2)^2, & u &= -\frac2{\sqrt{7}}\bigl(w+w^{-1}\bigr).
\end{align}
Thus, the potentials are the holomorphic functions on the Riemann surface of the function $w+w^{-1}$. The latter be constructed out of three copies of the complex plane with the cuts, going from the square root singularities at $x=\pm1$, where two roots of the initial cubic equation glue together. There are a contour depicted on Fig.~\ref{fig:riemann_cubic}, visiting all three sheets of the cover, such all three potentials under consideration are real along it, and coincide to the result, obtained form the parametric form.

\subsubsection*{Bicubic and biquartic cases}

There are the values of the parameters, namely $\ell=7$, $m = 4,\,2$ and $\ell=9$, $m = 6,\,4,\,2$, for which the equation $x = x(y)$ is also can be solved for $y$ or $u$ in quadratures, so the potential can be found explicitly. For $\ell=9$ the corresponding equation is bicubic, i.e.\ cubic equation on $y^2$ or $u^2$. Similarly, for $\ell=7$ we have quartic equation on $y^2$ or $u^2$, so it also can be solved explicitly. Nevertheless, it qualitatively similar to the cases, discussed above so we will not discuss it in detail.

\subsubsection*{Cases $(\ell,m) = (1,1),\;(2,2),\;(2,1),\;(3,2)$}

Finally, let us briefly discuss the values of the parameters (\ref{eq:pt_anal_pot_pars}), leading to the analytic potentials (\ref{eq:pt_anal_pots}). The cases of $\ell=2$,~$m=1$ and $\ell=3$,~$m=2$ lie in the class $m=\ell-1$, already described above. For $\ell=m=1$ we have the equation $x = x(y)$ which is linear on the parameter $v = \arcsin y$, so we have
\begin{equation}
	x(y) = 2\arcsin y = 2v, \qquad V_{1,1}(x) = \frac12(1-y^2) = 1+\cos x.
\end{equation}
This is the only case, corresponding to the potential, which is periodic after the continuation beyond the instanton image, and can be found explicitly. This is because for higher odd $\ell$ and $m$ the equation $x = x(y)$ becomes transcendental (cf.~(\ref{eq:x_of_y_parity})). Similarly, for $\ell=m=2$, the equation $x=x(y)$ is linear in $y$, so the potential is trivially reconstructed as
\begin{equation}
	x(y) = y, \qquad V_{2,2}(x) = \frac12(1-y^2)^2 = \frac12(1-x^2)^2.
\end{equation}

Thus, we constructed the generalization of well-known potentials (\ref{eq:pt_anal_pots}), demanding that its fluctuation operators on the instanton background are still P\"{o}schl-Teller operators (\ref{eq:PT_op}), but its parameters $\ell$ and $m$, parameterizing the number of bound state eigenfunctions and negative modes, differ from those of (\ref{eq:pt_anal_pot_pars}) and may be arbitrary. Then, using the technique developed in Section~\ref{sec:reconstruction} we obtained the infinite series of the potentials (\ref{eq:PT_pot_par}) parameterized by $\ell$ and $m$, and examined its properties.
	
\subsection{Isospectral deformations} \label{sec:isospectral}

	In this subsection, we consider one more generalization of mentioned potentials. Namely, we demand that the spectrum of the fluctuation operator of the generalized potential coincide to those of $V_{\ell,m}(x)$, including the known ones (\ref{eq:pt_anal_pots}), but allow the eigenfunctions to be different. This problem can be cast as so-called isospectral deformation of Schr\"{o}dinger operators \cite{Keung:1989vx}. 
	
	One of the most known family of the isospectral deformation problem is connected to Korteweg-de Vries (KdV) hierarchy of integrable equations. Specifically, the inverse scattering method for finding $\ell$-soliton solution KdV equation \cite{Gardner:1967wc,novikov1984theory}, for fixed integer $\ell$ and negative bound state eigenvalues $\tilde{\lambda}_j$,~$j=1,\ldots\ell$ gives $\ell$-parametric set of isospectral Schr\"{o}dinger operators $L_\ell$ together with the corresponding bound state and scattering state eigenvectors. These can be expressed via the matrix $A$, defined as
	\begin{equation}
		A_{ij}(\tau) = \delta_{ij}+\frac{\beta_i}{\varkappa_i+\varkappa_j} e^{-(\varkappa_i+\varkappa_j) \tau},
	\end{equation}
	in terms of which the operator reads
	\begin{equation} \label{eq:kdv_L_op}
		L_\ell = -\partial_\tau^2 + U_\ell(\tau), \qquad U_\ell(\tau) = -2 \partial_\tau^2 \ln\det A(\tau),
	\end{equation}
	where $\beta_k$,~$k=1,\ldots\ell$ is a set of positive parameters, defining a particular member of isospectral family. The operator has exactly $\ell$ bound state eigenvectors expressed as
	\begin{equation} \label{eq:kdv_eigenvector}
		\psi_j(\tau) = \frac{\det A^{(j)}(\tau)}{\det A(\tau)},
	\end{equation}
	and having the eigenvalues $\tilde{\lambda}_j = -\varkappa_j^2$, where the matrix $A^{(j)}$ is obtained from $A$ by replacing $j$-th column by $-\beta_i \, e^{\varkappa_i \tau}$,~$i=1,\ldots\ell$. Operators $L_\ell$ have no normalizable zero mode, since all bound state eigenfunctions $\psi_j$ correspond to negative eigenvalues. Nevertheless, one can assign each of these eigenfunctions the meaning of zero mode by shifting the operator, namely the operator $L_{\ell,j}:=L_{\ell} + \varkappa_j^2$ has the zero mode $\psi_j$. Now, one can straightforwardly apply the procedure of Section~\ref{sec:reconstruction} to $L_{\ell,j}$, thus obtaining $\ell$-parametric family of the potentials\footnote{In fact, it is $(\ell-1)$-parametric, since one-parametric transformation of the parameters $\beta_j \mapsto \beta_j e^{2\varkappa_j \tau_0}$, just makes an overall shift of the potential.}, whose fluctuation operators $L_{\ell,j}$ have the same spectra, defined by $\varkappa_i$,~$i=1,\ldots\ell$. 
	
	
	The operators $L_{\ell,m}=L_\ell+\varkappa_m^2$ coincide with P\"{o}schl-Teller operators $K_{\ell,m}$ defined in (\ref{eq:PT_op}), up to particular choice of isospectral parameters $\beta_j$ and spectrum-defining parameters $\varkappa_j$, namely
	\begin{align}
		\beta_j &= 2 \varkappa_j \prod_{i\ne j}\frac{\varkappa_i+\varkappa_j}{|\varkappa_i-\varkappa_j|}, \label{eq:beta_pt}\\
		\varkappa_j &= j, \label{eq:kappa_pt}
	\end{align} 
	so its eigenvalues has the form $\tilde{\lambda}_j + m^2 = m^2 - j^2$ as demanded (cf.\ (\ref{eq:PT_ev}) with $j = k-m$). Thus, P\"{o}schl-Teller operator $K_{\ell,m}$ is representative of whole the isospectral family, parameterized by $\beta_j$. This means that each potential $V_{\ell,m}(x)$, obtained in the previous section, admits $\ell$-parametric deformation that keeps fixed the spectrum of the fluctuation operator about the corresponding instanton.
	
\begin{figure}[h!]
		\centering
		\includegraphics[scale=1]{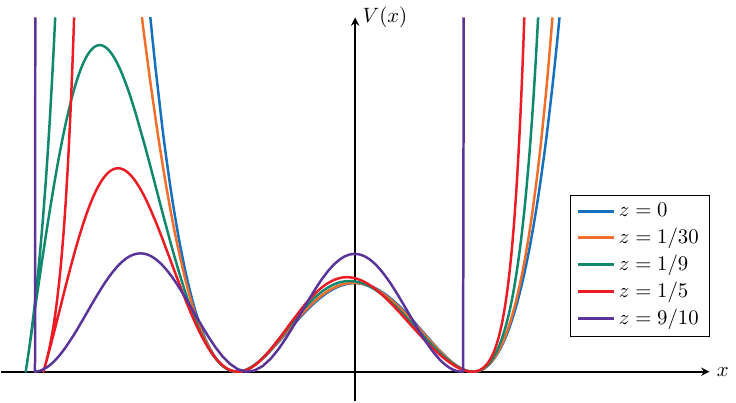}
		\caption{Reconstructed potential for the operator $L_{2,2}$ which is isospectral deformation of P\"{o}schl-Teller operator $K_{2,2}$ (corresponding to double-well potential) with deformation parameter~$z$.}
		\label{fig:dw_deformed}
	\end{figure}

	Let us explicitly construct such a deformation for $V_{2,2}(x)$, i.e.\ double-well potential. This is the simplest case admitting non-trivial deformation, since the deformation of sine-Gordon potential $V_{1,1}(x)$ only leads to overall shift and is not interesting. The corresponding fluctuation operator reads
	\begin{equation}
		L_{2,2} = -\partial_\tau^2 + U_2(\tau) - \varkappa_2^2,
	\end{equation}
	where $U_2(\tau)$ is obtained from (\ref{eq:kdv_L_op}) and have the form
	\begin{equation}
		U_2(\tau) = -\frac{2(\varkappa_2 ^2 -\varkappa_1^2)\bigl(\varkappa_2^2 \cosh^2\varkappa_1(\tau-\tau_1)+\varkappa_1^2 \cosh^2\varkappa_2(\tau-\tau_2)\bigr)}{\bigl(\varkappa_2\cosh\varkappa_2(\tau-\tau_2)\cosh\varkappa_1(\tau-\tau_1)-\varkappa_1\sinh\varkappa_2(\tau-\tau_2)\sinh\varkappa_1(\tau-\tau_1)\bigr)^2}.
	\end{equation}
	Here we introduce the parameters $\tau_{1,2}$ instead of the initial ones $\beta_{1,2}$ and connected to the latter as
	\begin{equation}
		\beta_{i} = 2 \varkappa_{i} \frac{\varkappa_1+\varkappa_2}{|\varkappa_1-\varkappa_2|} e^{2\varkappa_{i}\tau_{i}}, \qquad i=1,2,
	\end{equation}
	i.e.\ non-vanishing $\tau_{1,2}$ represent a measure of deviation from the P\"{o}schl-Teller case (\ref{eq:beta_pt}). Zero mode $\eta_0$ of $L_{2,2}$ coincide with eigenfunction $\psi_2$ of $L_{2}$, defined in (\ref{eq:kdv_eigenvector}), and has the following explicit form
	\begin{equation}
		\eta_0(\tau) = \frac{\cosh\varkappa_1(\tau-\tau_1)}{\varkappa_2\cosh\varkappa_2(\tau-\tau_2)\cosh\varkappa_1(\tau-\tau_1)-\varkappa_1\sinh\varkappa_2(\tau-\tau_2)\sinh\varkappa_1(\tau-\tau_1)}. 
	\end{equation}
	Thus, the only relevant parameter is $\tau_1-\tau_2$, since the simultaneous shift of $\tau_1$ and $\tau_2$ can be absorbed to the shift of $\tau$. Thus, we can set $\tau_1 = -\tau_2 = \tau_0$ without loss of generality. Substituting also the parameters $\varkappa_i$, defining the P\"{o}schl-Teller spectrum (\ref{eq:kappa_pt}), one obtains the following expression for $U_2$ and zero mode
	\begin{align}
		U_2(\tau) &= -12\frac{4\cosh2(\tau+\tau_0)+\cosh4(\tau-\tau_0)+3}{\bigl(3\cosh(\tau-3\tau_0)+\cosh(3\tau-\tau_0)\bigr)^2}, \\
		\eta_0(\tau) &= \frac{2\cosh(\tau-\tau_0)}{3\cosh(\tau-3\tau_0)+\cosh(3\tau-\tau_0)}.
	\end{align}
	It is easy to see that for $\tau_0=0$ the above quantities reduce to those of P\"{o}schl-Teller. Now, we can apply the reconstruction procedure, developed in Section~\ref{sec:reconstruction} to $\eta_0$. It is useful to not only express zero mode in terms of $y=\tanh\tau$ as in P\"{o}schl-Teller case, but also perform a similar substitution for $\tau_0$, namely $z=\tanh\tau_0$. Thus, zero mode reads
	\begin{align}
		\eta_0(\tau) &= R(\tanh \tau), \qquad R(y) = \frac{(1-z^2)(1-y^2)(zy-1)}{2z(2+z^2)y^3 + 6z^2y^2 - 6zy -2(2+z^2)}.
	\end{align}
	The integral (\ref{eq:x_of_tau}) can be done explicitly by using partial fraction decomposition, and the result is expressed in terms of distinct roots $y_{1,2,3}(z)$ of cubic polynomial in the denominator of $R(y)$. The underlying integral reads
	\begin{equation}
		x(y) =\nu \int^y \frac{dy'}{1-y'^2} R(y') = \nu \frac{1-z^2}{6z} \sum_{i=1}^3 \frac{(z y_i - 1)\ln(y-y_i)}{(z^2+2)y_i^2+2z y_i-1},
	\end{equation}
	where $\nu$ is normalization factor as before, so the instanton solution is $x_c(\tau) = x(\tanh\tau)$. The obtained deformed potential, parametrically defined as $(x, V(x)) = \bigl(x(y), \nu^2 R^2(y)/2\bigr)$, has the following properties, which can be derived similar to those of Appendix~\ref{sec:app_tp_PT}. The potential has two harmonic minima, corresponding to its counterparts in undefromed case. Additionally, at $x_0=x(1/z)$, lying beyond the instanton image, the potential has a zero, corresponding to zero $y = 1/z = \coth \tau_0$ of $R(y)$. At this point the potential has the square root branch point singularity, i.e.\ $V(x) = C_1 (x - x_0) (1 + \sqrt{C_2(x - x_0)}+\ldots)$ for some $z$-dependent $C_{1,2}$. Moreover, the $x^4$-like power-law growth of the potential for $|x|\to \infty$ in the undeformed case is replaced by the exponential growth in the deformed one. Typical shapes of the deformed potential for different values of the parameter is depicted at Fig.~\ref{fig:dw_deformed}.
	
	Thus the deformation of the double-well potential leads to additional minimum, where it has square root branch point singularity. For small values of the deformation parameter $z$ this point is far enough from the initial ones hence is not significant, but lies at finite distance for $z\to \pm 1$. Despite the significant modification of the potential shape, by the definition of the deformation procedure, its fluctuation operator has the same spectrum. Moreover if we set the normalization constant $\nu$ to be independent of deformation parameter, the instanton action also coincides with those of undeformed case. For instance, this means that one-instanton partition function is not sensitive to deformation at one loop.
	
	The deformation of the potentials $V_{\ell,m}(x)$ with higher values of the parameters can be done in the same way, but the technically much more laborious, hence we leave it beyond the scope of our discussion. Nevertheless, we note that in all higher cases the deformation leads to additional singularities of the potential.
	
%
\section{Natanzon and shape invariant operators} \label{sec:sip}

In Section~\ref{sec:pt_rec} we construct the potentials whose fluctuation operators on the instanton background are reflectionless P\"oschl-Teller operators. Latter operators lie in much wider class of so-called Natanzon operators \cite{natanzon1999study,Natanzon:1979sr}, which are the most general Schr\"{o}dinger operators whose eigenfunctions are expressible in terms of hypergeometric function. This multi-parameter class contains almost all known exactly solvable one-dimensional Schr\"{o}dinger operators, including so-called shape-invariant operators \cite{Dabrowska:1987fd,Cooper:1986tz,Cooper:1994eh} as its subclass, which can also be solved using the methods of supersymmetry.

The problem we faced in the case reflectionless P\"oschl-Teller operators is that the corresponding potentials reconstructed are non-analytic (except previously known cases (\ref{eq:pt_anal_pots})), that significantly complicates all underlying calculations, including path integral treatment, perturbation theory, etc. In this section we reconstruct potentials, having Natanzon operator as a fluctuation operator on the instanton background. Then, we try to fix the parameters such that the potential reconstructed is smooth along the real line\footnote{The Natanzon operators can also be constructed for the confluent hypergeometric function. However, using the arguments of Appendix~\ref{sec:app_analytic}, this case can be ruled out as giving necessarily non-analytic reconstructed potentials.}. Though we mostly focus on the subclass of shape-invariant operators, their treatment in terms of Natanzon ones allow us to account all them in a unified way.

\subsection{Overview and reconstruction}
Natanzon operator defines the eigenvalue problem
\begin{equation} \label{eq:nat_eq}
	\bigl[-\partial_\tau^2 + W_N(\tau)\bigr] \eta(\tau) = k^2 \eta(\tau)
\end{equation}
where $k^2$ is energy parameter, and $W_N(\tau)$ is called the Natanzon potential and defined parametrically as
\begin{equation} \label{eq:nat_W}
	W_N(\tau) = \frac{f z(z-1) + h_0 (1-z) + h_1 z + 1}{R(z)} + \left[a + \frac{a+(c_1-c_0)(2z-1)}{z(z-1)}-\frac54\frac{\Delta}{R(z)}\right]\frac{z^2(1-z)^2}{R^2(z)}
\end{equation}
where $R(z)$ is the quadratic polynomial
\begin{equation}
	R(z) = a z (z-1) + c_0 (1-z) + c_1 z
\end{equation}
and $\Delta = (a + c_0-c_1)^2 - 4 a c_0$ is its discriminant, whereas $z=z(\tau)$ is defined by the differential equation
\begin{equation} \label{eq:nat_z_eq}
	\frac{(\dot z)^{2}R(z)}{4z^2(1-z)^2} = 1.
\end{equation}
The solutions of (\ref{eq:nat_eq}) has the following form
\begin{equation} \label{eq:nat_ef}
	\eta(\tau)  = (\dot{z}(\tau))^{-1/2} (z(\tau))^{(\lambda_0+1)/2} (1-z(\tau))^{(\lambda_1+1)/2} F(z(\tau))
\end{equation}
where $F(z)$ is the solution of hypergeometric equation, i.e.\ a linear combination of its independent solutions e.g.\ ${}_2 F_1(\alpha,\beta;\gamma;z)$ and ${}_2F_1(\alpha,\beta;\alpha+\beta-\gamma+1;1-z)$, where $\alpha$,~$\beta$ and~$\gamma$ are expressible in terms of auxiliary parameters $\lambda_0$,~$\lambda_1$ and~$\mu$ defined by linear transformation
\begin{equation} \label{eq:nat_alpha_mu}
	\alpha = \frac12(\lambda_0+\lambda_1-\mu+1), \qquad
	\beta = \frac12(\lambda_0+\lambda_1+\mu+1), \qquad
	\gamma = \lambda_0+1,
\end{equation}
which has the following dependence on the energy parameter $k^2$
\begin{equation} \label{eq:nat_mu_k}
	1-\mu^2 = a k^2 - f, \qquad
	1-\lambda_p^2 = c_p k^2 - h_p, \quad p=0,1.
\end{equation} 
Normalizability of bound states selects the solutions $F(z)$ of hypergeometric equation in (\ref{eq:nat_ef}), which have no singularities either at $z=0$ and $z=1$ or at $z=\infty$ so $F(z) = {}_2F_1(\alpha,\beta;\gamma;z)$ with negative integer or zero $\alpha=-n$. Substituting it to (\ref{eq:nat_mu_k}) and solving for $\beta$ and $\gamma$, one obtains the quantization condition of the energy parameter
\begin{equation} \label{eq:nat_quantization}
	2n+1=\sqrt{f+1-a k_n^2}-\sqrt{h_0+1-c_0 k_n^2}-\sqrt{h_1+1-c_1 k_n^2}.
\end{equation}
As consequence, the parameters $\lambda_0$,~$\lambda_1$ and~$\mu$, defining the arguments of hypergeometric (\ref{eq:nat_alpha_mu}) of the hypergeometric function, are also quantized due to (\ref{eq:nat_mu_k}).

Thus, the Natanzon operators constitute the seven parametric family of Schr\"{o}dinger operators, where six parameters are $a,c_0,c_1,f,h_0,h_1$, and seventh parameter arises as integration constant of the equation (\ref{eq:nat_z_eq}). However, for our purposes it is useful to express part of them in terms of different parameters. Namely, instead of $h_0$,~$h_1$ and~$f$ we introduce new independent parameters $\bar{\lambda}_0$,~$\bar{\lambda}_1$ and $k_0^2$, where $k_0^2$ is the energy parameter of ground state, i.e.\ corresponding to $n=0$ in (\ref{eq:nat_quantization}), whereas $\bar{\lambda}_0$ and~$\bar{\lambda}_1$ equal to $\lambda_0$ and~$\lambda_1$ respectively, corresponding to $n=0$ in (\ref{eq:nat_mu_k}). Moreover, without loss of generality we set ground state eigenfunction to be zero mode, i.e.\ $k_0^2 = 0$. The explicit form of variables change follows from (\ref{eq:nat_mu_k}),~(\ref{eq:nat_quantization}) and reads
\begin{equation} \label{eq:nat_lam_par}
	f = (\bar{\lambda}_0 + \bar{\lambda}_1+1)^2 - 1, \qquad h_p = \bar{\lambda}_p^2-1, \quad p=0,1.
\end{equation}

Now, we are ready to apply reconstruction procedure described in Section~\ref{sec:reconstruction}. Once we are looking for analytic reconstructed potentials, according to the analysis of Appendix~\ref{sec:app_tp_general} we should only use the ground state eigenfunction and, possibly, first exited state eigenfunction if the latter is odd, as a zero mode. Since the Natanzon operators generally have no parity we will focus on the ground state case.
Using that ground state correspond to $\alpha=-n=0$, so $F(z)={}_2F_1(0,\beta;\gamma;z)=1$, we find that the ground state eigenfunction is simply $\eta_0(\tau) = \dot{z}^{-1/2} z^{(\bar{\lambda}_0+1)/2} (1-z)^{(\bar{\lambda}_1+1)/2}$. Thus, using (\ref{eq:pot_rec_par}) we obtain the parametric form of the desired potential $V_N(x)$ as $(x, V_N(x)) = (x(z), V_N(x(z)))$, where
\begin{align}
	&x(z) = \nu \int^{\tau(z)} d\tau \, \eta_0(\tau)
	= 2^{-3/2} \,\nu \int dz \, \bigl(R(z)\bigr)^{3/4} \, z^{\bar{\lambda}_0/2-1} (1-z)^{\bar{\lambda}_1/2-1}, \label{eq:nat_x_z}\\
	&V_N(x(z)) = \frac12 \nu^2 \bigl[\eta_0(\tau(z))\bigr]^2 = \frac14 \nu^2 \sqrt{R(z)} \, z^{\bar{\lambda}_0} (1-z)^{\bar{\lambda}_1}, \label{eq:nat_V_z}
\end{align}
and we explicitly substitute $\dot{z}(\tau)$ expressed from (\ref{eq:nat_z_eq}), whereas $\nu$ is normalization factor. Equations (\ref{eq:nat_x_z}) and~(\ref{eq:nat_V_z}) formally solve the problem of finding potential $V_N(x)$, having Natanzon operator as a fluctuation operator. The next task is to select the values of the Natanzon operator parameters, such that the potential reconstructed is smooth along the real $x$-line.

\subsection{Shape invariant operators and analyticity}

Although the explicit form of $\tau(z)$ can be found from the equation (\ref{eq:nat_z_eq}) for arbitrary values of the parameters of Natanzon operator, the explicit form of the inverse function $z(\tau)$ can be obtained only for specific values of the parameters, namely $a,c_0,c_1$ should be such that $R(z)$ has roots only at $z=0$ and/or $z=1$. This choice of $R(z)$ corresponds exactly to six well-known exactly solvable operators, which constitute subclass of shape-invariant operators in the class of Natanzon operators \cite{Cooper:1994eh,Derezinski:2010ku}, namely trigonometric P\"{o}schl-Teller, hyperbolic P\"{o}schl-Teller, Manning-Rosen, Eckart, Scarf and Rosen-Morse potentials. We will concentrate on this subclass in finding Natanzon operators leading smooth reconstructed potentials. Corresponding choices of $R(z)$ and appropriate solutions to differential equation (\ref{eq:nat_z_eq}) are summarized in the table
\begin{equation} \label{eq:nat_pot_table}
	\text{
	\begin{tabular}{|c|c|c|c|} \hline
		& $z(\tau)$ & $\dot{z}$&$R(z)$ \\\hline
		MR & $(1+e^{2\tau})^{-1}$& $2z(z-1)$ & $1$\\\hline
		hPT & $\tanh^2\tau$ & $2z^{1/2}(1-z)$& $z$\\\hline
		E & $1-e^{-2\tau}$& $2(1-z)$& $z^2$\\\hline
		tPT & $\sin^2 \tau$& $2z^{1/2}(1-z)^{1/2}$& $z(1-z)$\\\hline
		S & $\tfrac12(1-i\sinh \tau)$ & $ (-z)^{1/2} (1-z)^{1/2}$ & $-4z(1-z)$ \\\hline
		RM & $\tfrac12(1+i\cot \tau)$& $-2iz(1-z)$& $-1$\\\hline
	\end{tabular}}
\end{equation}
The explicit form of $R(z)$ fixes the parameters $a,c_0,c_1$, the explicit form of $z(\tau)$ defines the integration constant, whereas $f,h_0,h_1$ are the free parameters (equivalently, $\bar{\lambda}_0,\bar{\lambda}_1, k_0^2$). 

To handle all these operators in the unified way, we parameterize $R(z)$ as
\begin{equation}
	R(z) = A \, z^{q_0} (1-z)^{q_1}, 
\end{equation}
where $q_0,q_1 = 0,1,2$ and $q_0+q_1 \le 2$. Introducing parameters $r_0$,~$r_1$ instead of $\bar{\lambda}_0$,~$\bar{\lambda}_1$ as
\begin{equation} \label{eq:nat_r_par}
	r_p = \frac12\bar{\lambda}_p + \frac34 q_p, \qquad p=0,1
\end{equation}
we rewrite the parametric form of reconstructed potential as
\begin{align} \label{eq:nat_sip_pot}
	x(z) = \tilde{\nu} \int dz \, z^{r_0-1} &(1-z)^{r_1-1}, \qquad V_N(x(z)) =  \frac{2\tilde{\nu}^2}{A} \, z^{2r_0-q_0} (1-z)^{2r_1-q_1}
\end{align}
where $\tilde{\nu} = 2^{-3/2} A^{3/4} \nu$. The function $x(z)$ defining the instanton trajectory is, in fact, incomplete beta-function. It is useful to express it in terms of hypergeometric function, and using the properties of the latter, write down in three different forms, corresponding to the series expansion near $z=0$, $z=1$ and $z=\infty$, respectively
\begin{subequations} \label{eq:nat_inst_traj}
\begin{align}
	x(z) &= \phantom{-}\frac{\tilde{\nu}}{r_0}z^{r_0} {}_2F_1(r_0, 1-r_1,r_0+1;z) + x_0,  \label{eq:nat_inst_traj_0}\\
	&=-\frac{\tilde{\nu}}{r_1}(1-z)^{r_1} {}_2F_1(r_1, 1-r_0,r_1+1;1-z) + x_1, \label{eq:nat_inst_traj_1}	\\
	&= -\frac{\tilde{\nu}\,e^{\pm i\pi r_1}}{1-r_0-r_1} z^{-(1-r_0-r_1)}{}_2F_1(1-r_1,1-r_0-r_1,2-r_0-r_1,z^{-1}) + x_\infty, \label{eq:nat_inst_traj_infty}
\end{align}
\end{subequations}
where $x_0$,~$x_1$, and $x_\infty$ are integration constants.
In Appendix~\ref{sec:app_tp_nat} we show that near turning points $x_{\mathrm{tp}}$, corresponding to $z=0$, $z=1$ and $z=\infty$, the inverse function $z(x)$ has the form of the series in powers of $(x-x_{\mathrm{tp}})^{1/r_0}$, $(x-x_{\mathrm{tp}})^{1/r_1}$, and $(x-x_{\mathrm{tp}})^{1/(1-r_0-r_1)}$, respectively. Substitution of the inverted function to the parametric form of the potential shows, that the potential also has the form of such series. Hence, the restrictions on the parameters $r_0$,~$r_1$ come from the requirement that the these powers are integer positive numbers.

For the first four operators in (\ref{eq:nat_pot_table}) the function $z(\tau)$ ranges from $z=0$ to $z=1$, and the same points correspond to turning points of the reconstructed potentials. At the same time, for the last two operators in the same table the range of $z(\tau)$ is the line, parallel to imaginary axis, passing through $z=1/2$, so both turning points correspond to $z=\infty$. Thus, according to the discussion of Appendix~\ref{sec:app_tp_nat} the condition of analyticity near the turning points for the first four operators read 
\begin{equation} \label{eq:nat_anal_01}
	r_0 = \frac1{N_0}, \qquad r_1 = \frac1{N_1},
\end{equation}
while for the last two operators the condition is 
\begin{equation} \label{eq:nat_anal_infty}
	r_0 + r_1 = 1 - \frac1M,
\end{equation}	
where $N_0$,~$N_1$, and~$M$ are positive integers. 
In addition, the finiteness of the zero mode, which can be written as $\eta_0(\tau(z)) = (A/4)^{1/4}\, z^{r_0-q_0/2}(1-z)^{r_1-q_1/2}$ near $z=0$, $z=1$, and $z=\infty$ requires
\begin{equation}
	2r_0 \ge q_0, \qquad
	2r_1 \ge q_1, \qquad
	2(r_0+r_1) \le q_0 + q_1, \label{eq:nat_eta_reg}
\end{equation}
respectively.

Trigonometric P\"{o}schl-Teller, Eckart and Rosen-Morse operators necessarily have $1/\tau^2$-singulari\-ties, so these ones have no chance to lead to smooth reconstructed potentials for the reasons presented in Appendix~\ref{sec:app_tp_types}. Thus, we will consider only hyperbolic P\"{o}schl-Teller, Manning-Rosen and Scarf potentials. For each of these three operators we will deduce the range of parameters $r_0$,~$r_1$ leading to analytic reconstructed potential, ensuring that zero mode $\eta_0(\tau)$ is bounded and normalizable. Then, we will write down corresponding parametric form (\ref{eq:nat_sip_pot}) of the reconstructed potential, classify the cases in which the function $x(z)$ can be inverted explicitly, thus giving the explicit form of the potential, and describe the procedure of continuation beyond the instanton image.

\subsubsection*{Manning-Rosen operator}

\begin{figure}[h!]
	\centering
	\begin{subfigure}{0.49\textwidth}
		\centering
		\includegraphics[scale=0.9]{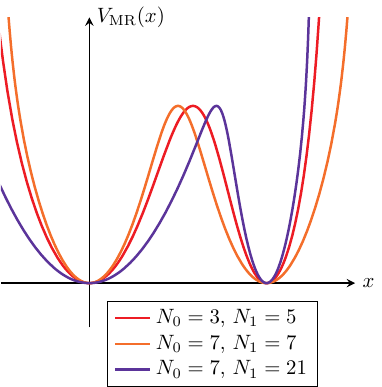}
		\caption{}
	\end{subfigure}	
	\begin{subfigure}{0.49\textwidth}
		\centering
		\includegraphics[scale=0.9]{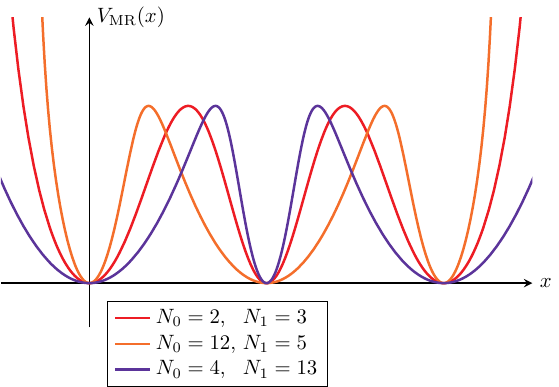}
		\caption{}
	\end{subfigure} \\
	\begin{subfigure}{0.49\textwidth}
		\centering
		\includegraphics[scale=0.9]{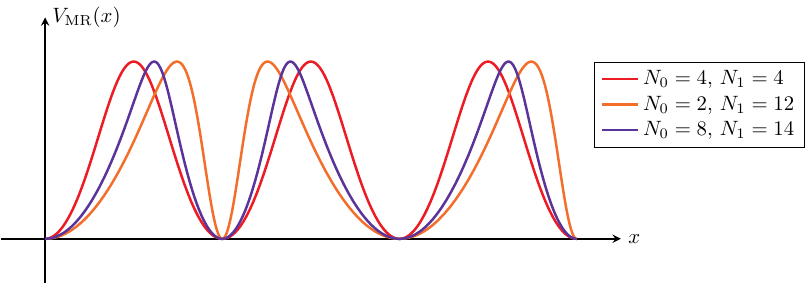}
		\caption{}
	\end{subfigure}
	\caption{Analytic reconstructed potentials (\ref{eq:nat_pot_MR}), whose fluctuation operators are Manning-Rosen operators. Cases (i)--(iii) correspond to different parities of parameters $N_0$,~$N_1$.}
	\label{fig:nat_MR}
\end{figure}

For Manning-Rosen operator, the function $z(\tau)=1/(1+e^{2\tau})$ connects $z=0$ and $z=1$ so the restrictions on $r_0$, $r_1$ are given by (\ref{eq:nat_anal_01}). There are no additional restrictions on $r_0 = 1/N_0$ and $r_1 = 1/N_1$ coming from the regularity condition (\ref{eq:nat_eta_reg}).

Subsequent substitution of $r_0$, $r_1$ to (\ref{eq:nat_r_par}), (\ref{eq:nat_lam_par}) and (\ref{eq:nat_W}) gives the function $W(\tau)$, parameterizing Manning-Rosen operator
\begin{align}
	W_{\text{MR}}(\tau) = -\frac{\ell_+(\ell_+ +1)}{\cosh^2 \tau} + 2 \,\ell_+ \ell_- \tanh \tau + \ell_+^2 + \ell_-^2, \qquad \ell_\pm = \frac1{N_0} \pm \frac1{N_1}
\end{align}
For the particular case $N_0=N_1$ the second term vanishes, so $W_{\text{MR}}(\tau)$ reduces to those of P\"{o}schl-Teller operator with fractional $\ell = 2/N_0$, and correspond to the double-well and sine-Gordon potential for $N_0=1,2$, respectively.

We choose the normalization $\tilde\nu = -\Gamma(r_0+r_1)/\Gamma(r_0)\Gamma(r_1)$ and integration constant in (\ref{eq:nat_sip_pot}) such that $x(0)=1$, $x(1)=0$. Thus, the instanton trajectory near $z=0$ and $z=1$ has the forms
\begin{subequations}
	\begin{align}
		x(z) &= -N_0\frac{\Gamma(\frac1{N_0}+\frac1{N_1})}{\Gamma(\frac1{N_0})\Gamma(\frac1{N_1})} z^{\frac1{N_0}}{}_2F_1(\tfrac1{N_0},1-\tfrac1{N_1};1+\tfrac1{N_0};z)+1 \label{eq:nat_MR_x_0}\\&= 
		N_1\frac{\Gamma(\frac1{N_0}+\frac1{N_1})}{\Gamma(\frac1{N_0})\Gamma(\frac1{N_1})} (1-z)^{\frac1{N_1}}{}_2F_1(\tfrac1{N_1},1-\tfrac1{N_0};1+\tfrac1{N_1};1-z), \label{eq:nat_MR_x_1}
	\end{align}
\end{subequations}
respectively, whereas the potential reads
\begin{equation} \label{eq:nat_pot_MR}
	V_{\text{MR}}(x(z)) = 2 \left[\frac{\Gamma(\frac1{N_0}+\frac1{N_1})}{\Gamma(\frac1{N_0})\Gamma(\frac1{N_1})}\right]^2 z^{\frac2{N_0}} (1-z)^{\frac2{N_1}}
\end{equation}
For $N_0=1$ (or $N_1=1$) the hypergeometric function in (\ref{eq:nat_MR_x_0}),~(\ref{eq:nat_MR_x_1}) degenerates to the power-law function $x(z) = (1-z)^{1/N_1}$, so inverting it explicitly, we obtain the potential
\begin{equation} \label{eq:nat_MR_expl}
	V_{\text{MR}}(x) = \frac{2}{N_1^2} \, x^2 \bigl(1- x^{N_1}\bigr)^2, \qquad N_0 = 1,
\end{equation}
which has the shape of generally non-symmetric double-well for odd $N_1$ and symmetric triple-well for even $N_1$.

For general $N_0$,~$N_1$ the analytic continuation near $z=0$ can be done by the substitution $z=u^{N_0}$ (i.e.\ replacing $z^{1/N_0}$ by $u$) in (\ref{eq:nat_MR_x_0}) with  $-\infty < u < 1$ for odd $N_0$ and $-1 < u < 1$ for even $N_0$. Similarly, near $z=1$ analytic continuation can be done by substitution $z=1-v^{N_1}$ in (\ref{eq:nat_MR_x_1}) with  $-\infty < v < 1$ for odd $N_1$ and $-1 < v < 1$ for even $N_1$. The difference between odd and even case comes from the fact that in the even case $u=-1$ and $v=-1$ correspond to the singularity of hypergeometric function. Since both instanton trajectory and the potential are symmetric under the simultaneous change $N_0 \leftrightarrow N_1$, $z \leftrightarrow 1-z$, we should separately consider analytic continuation in three cases, corresponding to different parities of $N_0$ and $N_1$.
\begin{enumerate}[label=(\roman*)]
	\item \emph{$N_0$ --- odd, $N_1$ --- odd.} This is the simplest case in which $u$-parameterization covers $0 < x < + \infty$, whereas $v$-parameterization covers $-\infty < x < 1$, thus giving the values of  $V_{\text{MR}}(x)$ for all real $x$. Thus, $V_{\text{MR}}(x)$ has the shape of generally non-symmetric double-well potential with degenerate minima at $x=0,1$.
	\item \emph{$N_0$ --- even, $N_1$ --- odd.} In this case $v$-parameterization covers $-\infty < x < 1$, but $u$-parameterization covers only $0 < x < 2$. However, since $N_0$ is even, in this parameterization $x(z)-1$ is odd function of $u$, whereas the potential is even function of $u$. Thus, the potential is symmetric about $x=1$, hence $1 < x < + \infty$ can be covered, by replacing $x(z)$ by $2-x(z)$, where $x(z)$ is taken in $v$-parameterization. Thus,  $V_{\text{MR}}(x)$ has the shape of symmetric triple-well potential, with the degenerate minima at $x=0,1,2$.
	\item \emph{$N_0$ --- even, $N_1$ --- even.} For these parities $u$- and $v$-parameterizations cover finite intervals $0 < x < 2$ and $-1 < x < 1$, respectively. Nevertheless, $x(z)-1$ is odd function of $u$, instanton trajectory $x(z)$ is odd function of $v$, whereas $V_{\text{MR}}(x)$ is even function of both $u$ and $v$. This means that the potential is symmetric about both $x=0$ and $x=1$, hence it is periodic with the period 2, and degenerate minima at integer $x$.
\end{enumerate}
Plots of typical shapes of these potentials are depicted at Fig.~\ref{fig:nat_MR}.

\subsubsection*{Hyperbolic P\"{o}schl-Teller operator}

\begin{figure}[h!]
	\centering
	\begin{subfigure}[c]{0.49\textwidth}
		\centering
		\includegraphics[scale=0.9]{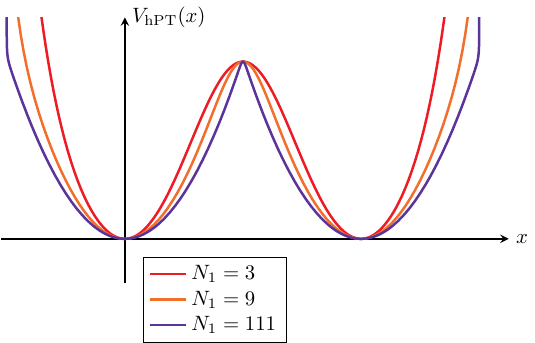}
		\caption{}
	\end{subfigure}	
	\begin{subfigure}[c]{0.49\textwidth}
		\centering
		\includegraphics[scale=0.9]{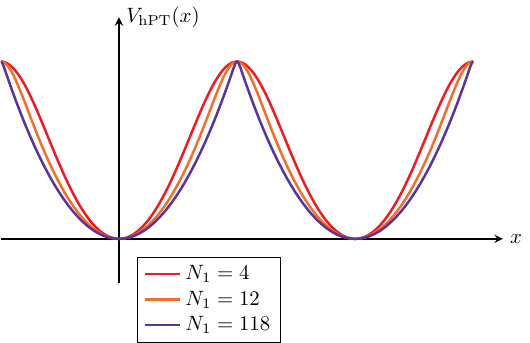}
		\caption{}
	\end{subfigure}
	\caption{Analytic reconstructed potentials (\ref{eq:nat_pot_hPT}), whose fluctuation operators are hyperbolic P\"{o}schl-Teller operators. Cases (i)--(ii) correspond to different parities of parameter~$N_1$ and fixed $N_0=2$.}
	\label{fig:nat_hPT}
\end{figure}

For hyperbolic P\"{o}schl-Teller operator, the function $z(\tau)=\tanh^2 \tau$ connects $z=0$ and $z=1$ so the restrictions on $r_0$, $r_1$ are given (\ref{eq:nat_anal_01}).
Normalizability near $z=1$ does not lead to additional condition on $r_1=1/N_1$ whereas regularity at $z=0$ requires $N_0=1,2$.

Subsequent substitution of $r_0$, $r_1$ to (\ref{eq:nat_r_par}), (\ref{eq:nat_lam_par}) and (\ref{eq:nat_W}) gives the function $W(\tau)$, parameterizing underlying operator
\begin{align}
	W_{\text{hPT}}(\tau) = -\frac{\ell(\ell+1)}{\cosh^2 \tau} + m^2, \qquad \ell = \frac2{N_0} + \frac2{N_1} -1, \quad m = \frac2{N_1}
\end{align}
where, generally, there is additional singular term $(\frac2{N_0} - 2)(\frac2{N_0}-1) /{\sinh^2 \tau}$, vanishing for $N_0 = 1,2$. In contrast to the reflectionless P\"{o}schl-Teller operators of Section~\ref{sec:pt_rec} the parameters $\ell$ and $m$ are not integer, but generally rational numbers of the special form. Looking for the particular values of $N_0$,~$N_1$ leading to integer $\ell$,~$m$, we deduce that $N_0=2$,~$N_1=2$ case corresponds to sine-Gordon potential, the values $N_0=2$,~$N_1=1$ correspond to double-well potential, $N_0=1$,~$N_1=1$ lead to cubic-well potential, whereas $N_0=1$,~$N_1=2$ correspond to inverted double-well potential.

The instanton trajectory coincides to those (\ref{eq:nat_MR_x_0}),~(\ref{eq:nat_MR_x_1}) of Manning-Rosen one, but the potential is different and reads
\begin{equation} \label{eq:nat_pot_hPT}
	V_{\text{hPT}}(x(z)) = 2 \left[\frac{\Gamma(\frac12+\frac1{N_1})}{\Gamma(\frac12)\Gamma(\frac1{N_1})}\right]^2 z^{\frac2{N_0}-1}(1-z)^{\frac2{N_1}}.
\end{equation}
As we seen in Manning-Rosen case, for $N_0=1$ the hypergeometric function in (\ref{eq:nat_MR_x_0}) degenerates to the power-law function $x(z) = (1-z)^{1/N_1}$, so it can be inverted explicitly, so the corresponding potential reads
\begin{equation} \label{eq:nat_hPT_expl}
	V_{\text{hPT}}(x) = \frac{2}{N_1^2} \, x^2 \bigl(1- x^{N_1}\bigr), \qquad N_0=1,
\end{equation}
and have the shape of inverted double-well potential for even $N_1$ and cubic-well potential for odd $N_1$.

For $N_0=2$ instanton trajectory degenerates to the elementary functions only for $N_1=1,2$, corresponding to double-well and sine-Gordon potentials, respectively.
For general $N_1$, analytic continuation near $z=0$ can be done by the substitution $z=u^2$ in (\ref{eq:nat_MR_x_0}) with $N_0=2$ and $-1 < u < 1$. In this parameterization $x(z)-1$ is odd function of $u$, whereas the potential $V_{\text{hPT}}(x)$ is even function of $u$, hence the potential is symmetric about $x=1$.
Near $z=1$ analytic continuation can be done by substitution $z=1-v^{N_1}$ in (\ref{eq:nat_MR_x_1}) with  $-\infty < v < 1$ for odd $N_1$ and $-1 < v < 1$ for even $N_1$, hence the continuation procedure is different for different parities.
\begin{enumerate}[label=(\roman*)]
	\item \emph{$N_1$ --- odd.} For this parity, $v$-parameterization covers $-\infty < x < 1$. Since the potential is symmetric about $x=1$, the interval $0 < x < + \infty$ can be covered by using $2-x(z)$ instead of $x(z)$ in $v$-parameterization. Thus, the potential $V_{\text{hPT}}(x)$ has the shape of symmetric double-well potential with the minima at $x=0,2$.	
	\item \emph{$N_1$ --- even.}	In this case $v$-parameterization covers $-1 < x < 1$. Instanton trajectory $x(z)$ is odd function of $v$, whereas $V_{\text{hPT}}(x)$ is even function of $v$. This means that the potential is symmetric about $x=0$. Thus, the $V_{\text{hPT}}(x)$ can be periodically continued, and its minima lie at even integer $x$.
\end{enumerate}
Plots of typical shapes of these potentials are represented at Fig.~\ref{fig:nat_hPT}.

\subsubsection*{Scarf operator}

\begin{figure}[h!]
	\centering
	\begin{subfigure}[c]{0.49\textwidth}
		\centering
		\includegraphics[scale=0.9]{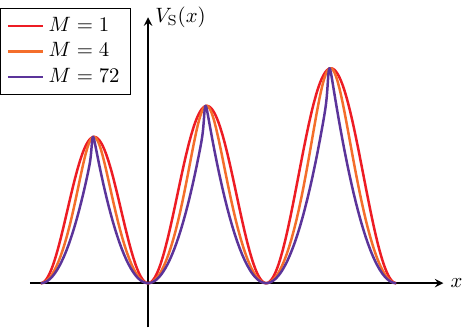}
		\caption{}
	\end{subfigure}	
	\begin{subfigure}[c]{0.49\textwidth}
		\centering
		\includegraphics[scale=0.9]{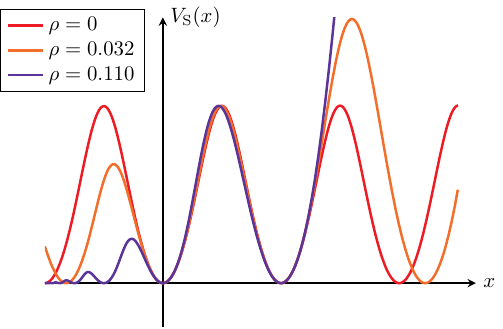}
		\caption{}
	\end{subfigure}
	\caption{Analytic reconstructed potentials (\ref{eq:nat_S_pot}), whose fluctuation operators are Scarf operators. (i) fixed $\rho=0.015$, different $M$, (ii) fixed $M=2$, different $\rho$.}
	\label{fig:nat_S}
\end{figure}

For Scarf operator, the function $z(\tau) = \frac12(1- i \sinh \tau)$ connects $z=\infty$ to itself, going through $z=1/2$ parallel to imaginary axis. Thus, we have a single restriction on $r_0$,~$r_1$ from the analiticity of the potential at turning points, namely (\ref{eq:nat_anal_infty}). However the function $W(\tau)$ defining the Scarf operator is real only when $r_0$ and $r_1$ are complex conjugate to each other. Thus, we parameterize it as
\begin{equation}
	r_0 = \frac12\left(1-\frac1{M}\right) + i \rho, \qquad r_1 = \frac12\left(1-\frac1{M}\right) - i \rho,
\end{equation}
where $\rho$ is an arbitrary real parameter.
\begin{equation}
	W_{\text{S}}(\tau) = - \frac{\ell(\ell+1)-4\rho^2}{\cosh^2\tau} - 2\rho \bigl(2\ell+1\bigr)\frac{\sinh \tau}{\cosh^2 \tau} + \ell^2, \qquad
	\ell = \frac1M
\end{equation}
The convenient parameterization of the instanton trajectory for Scarf case is (\ref{eq:nat_inst_traj_infty}) corresponding to expansion near $z = \infty$ and reads
\begin{subnumcases}{\label{eq:nat_S_x}x(z) =}
	- e^{-\frac{i\pi}{2M}+\pi\rho} G(z) + 1 \label{eq:nat_S_x_0}\\ 
	\phantom{-}e^{\frac{i\pi}{2M}-\pi\rho} G(z) \label{eq:nat_S_x_1}
\end{subnumcases}
where we introduce the function $G(z)$ as
\begin{equation} \label{eq:nat_S_G}
	G(z) = \frac1{2\pi}\frac{\Gamma(\frac12{+}\frac1{2M}{-}i\rho)\Gamma(\frac12{+}\frac1{2M}{+}i\rho)}{\Gamma(1+\frac1{M})} z^{-\frac1M} {}_2F_1(\tfrac1M,\tfrac12{+}\tfrac1{2M}{+}i\rho;1{+}\tfrac1M;z^{-1}).
\end{equation}
The first line in (\ref{eq:nat_S_x}) is valid for $\Im z < 0$, while the second line takes place for $\Im z > 0$. Change of the imaginary part sign correspond to the crossing of the cut of the hypergeometric function at $z^{-1}=2$. Hence, the second line can be deduced from the first one using the monodromy property of the hypergeometric function about $z=1$ (see e.g.\ \cite[Section~2.7.1]{erdelyi1953higher}). The normalization factor and integration constants are chosen such that $x(\frac12-i\infty) = 1$, $x(\frac12+i\infty) = 0$. The parametric expression for the potential read
\begin{equation} \label{eq:nat_S_pot}
	V_{\text{S}}(x(z)) = \frac1{8\pi^2}\left[ \frac{\Gamma(\frac12{+}\frac1{2M}{-}i\rho)\Gamma(\frac12{+}\frac1{2M}{+}i\rho)}{\Gamma(\frac1M)} \right]^2 \, z^{-\frac1M+2i\rho} \, (1-z)^{-\frac1M-2i\rho}.
\end{equation}

The instanton trajectory $x(z)$ reduces to the elementary function only for $M=1$, namely $x(z) = [(z^{-1}-1)^{-i\rho}-e^{-\pi\rho}]/2\sinh \pi\rho$, so it can be inverted explicitly and substituted to the parametric expression for the potential. The result reads
\begin{equation} \label{eq:nat_S_expl}
	V_{\text{S}}(x) = \frac{\rho^2}{2\sinh^2\pi\rho} \bigl(e^{-\pi\rho}(1-x)+e^{\pi\rho}x\bigr)^2\cos^2\left[\tfrac1{2\rho}\log(e^{-\pi\rho}(1-x)+e^{\pi\rho}x)\right], \qquad M=1.
\end{equation}
For $\rho\ne0$ this potential is not periodic, but rather quasi-periodic, in the sense that
\begin{equation} \label{eq:nat_S_shift}
	V_{\text{S}}(x+1) = e^{4\pi\rho} \, V_{\text{S}}(e^{-2\pi\rho}x).
\end{equation}
Now, analytically continuing the potential (\ref{eq:nat_S_pot}) to whole real $x$-line, we will show that this property takes place for general $M$.

For the purposes of analytic continuation of the potential beyond the instanton image, it is useful to parameterize $z$ as $z=1/(1+e^{-iy}) \equiv \frac12(1+i\tan \frac{y}2)$,~$-\pi < y < \pi$, such that $z$ has fixed real part equal to $1/2$ and arbitrary imaginary part, as needed. Thus, (\ref{eq:nat_S_x_0}) is valid for $-\pi < y < 0$, whereas one should use (\ref{eq:nat_S_x_1}) for $0 < y < \pi$ and cut crossing occurs for $y=0$. Analytic continuation can be done by extending the range of $y$ to arbitrary real values. For instance, extension to $\pi < y < 2\pi$ can be obtained by substitution $z^{-1/M} = 2^{1/M} e^{i u^M/M - i\pi/2M}S_M(u)$, $-(\pi/2)^{1/M} < u < (\pi/2)^{1/M}$, where $S_M(u)$ is the analytic function coinciding with $(\sin u^M)^{1/M}$ for non-negative real $u$. The substitution deduced from the requirement that $z^{-1} = 1 - e^{2iu^M}$ and $z^{-1/M}$ is an analytic function of $u$. Comparing the phase of $z^{-1/M}$ for small positive and negative $u$, we conclude that the range $\pi < y < 2\pi$ can be covered by $x(z)$ given by (\ref{eq:nat_S_x_1}), multiplied by the additional phase factor $e^{i\pi(1-1/M)}$. Extension to the range $2\pi < y < 3\pi$ correspond to cut crossing at $y=2\pi$ and can be done using monodromy property of hypergeometric function as in the case of $y=0$. Proceeding in the such way, we deduce the general formula for the instanton trajectory
\begin{equation} \label{eq:nat_S_x_y}
	x(z) = \frac{1-e^{-2k\pi \rho}}{1-e^{2\pi\rho}} + e^{-(2k+1)\pi\rho} \, G(z) \times \left\{
		\begin{aligned}
			&e^{\frac{i\pi}{2M}}, & 2\pi k &< y < \pi(2k+1), \\ -&e^{\frac{-i\pi}{2M}},& \pi(2k+1) &< y <\pi (2k+2),
		\end{aligned}
	\right.
\end{equation}
where $z = 1/(1+e^{-iy})$, $G(z)$ is defined in (\ref{eq:nat_S_G}), and $k$ is arbitrary integer. Performing $y$-substitution to (\ref{eq:nat_S_pot}), we obtain
\begin{equation} \label{eq:nat_S_pot_y}
	V_{\text{S}}(x(z)) = \frac1{8\pi^2} \left[ \frac{\Gamma(\frac12{+}\frac1{2M}{-}i\rho)\Gamma(\frac12{+}\frac1{2M}{+}i\rho)}{\Gamma(\frac1M)} \right]^2 2^{\frac2M} \left(\cos^2\tfrac{y}2\right)^{\!\frac1M} e^{-2\rho y}.
\end{equation}
Now, the property (\ref{eq:nat_S_shift}) can be easily checked, using (\ref{eq:nat_S_x_y}) and (\ref{eq:nat_S_pot_y}). Thus the potential $V_{\text{S}}(x)$ has the shape of the initial piece corresponding to $0 < x <1$, copied infinitely many times, and scaled such that neighboring pieces differ by $e^{4\pi \rho}$ in magnitude and by $e^{2\pi \rho}$ in width. The degenerate minima are placed at $x_k = (1-e^{-2k\pi \rho})/(1-e^{2\pi\rho})$ and accumulating at $x_{k=\infty}=1/(1-e^{2\pi\rho})$. Though the potential has single non-analyticity at the latter point, it is still analytic on the range of each particular instanton trajectory. The typical shapes of the potential for different parameters are shown at Fig.~\ref{fig:nat_S}.

\section{Discussion}\label{sec:discussion}

In this paper, we have developed the technique, described in Section~\ref{sec:reconstruction}, allowing to reconstruct the potentials of quantum-mechanical systems exhibiting instanton effects from known fluctuation operator, describing the linear perturbations of the instanton background.
This operator, together with its spectrum and the Green's function, constitutes clue elements of the perturbation theory on the instanton background. A particular practical interest of these construction is that it provides the way to generate infinite number of the systems, where the aforementioned ingredients are known explicitly, basically by starting not from a particular quantum-mechanical itself, but from the exactly solvable Schr\"{o}dinger operator, i.e.\ those for which the spectrum and the Green's function are known a priori. 

The basic idea, which allows us to succeed in developing the aforementioned technique is that the fluctuation operator, defining the quadratic action of the theory, on non-trivial background always has the zero mode, corresponding to the Euclidean time translation invariance. This mode is simply the Euclidean time derivative of the instanton solution. Thus, integrating the zero mode of the fluctuation operator, we obtain the instanton solution. Moreover, this derivative also defines the kinetic energy of the Euclidean equations of motion, which equals to the potential one on the instanton solution. This allows us to write down the parametric expression (\ref{eq:pot_rec_par}) for the potential, with the Euclidean time as the parameter.

Most known potentials, exhibiting tunneling (i.e.\ instanton effects), namely double-well, sine-Gordon and cubic well potentials, the fluctuation operators belong to the two-parametric family of reflectionless P\"{o}schl-Teller operators and correspond to the specific values of the parameters. Thus, in Section~\ref{sec:pt_rec} we try to generalize the aforementioned potentials, applying the procedure of reconstruction to the reflectionless P\"{o}schl-Teller operators for arbitrary values of its parameters. We found that for all choices of these parameters, despite the corresponding to the previously known potentials, the reconstruction procedure lead to the non-analytic potentials, which can be naturally defined on appropriately chosen Riemann surfaces. 

Obtained non-analytic potentials are real on the special contours on these Riemann surfaces, constructed for several specific cases in Section~\ref{sec:examples}, so in order to quantize the underlying systems, one can apply the quantization scheme similar to those used in the context of PT symmetric quantum mechanics \cite{Bender:1998ke,Bender:2007nj}.
Moreover, for non-analytic potentials the surface of the constant energy correspond to higher genus Riemann surfaces, so the corresponding instantons have richer classification by fundamental cycles, than in double-well case \cite{Nekrasov:2018pqq}.

Non-analyticity of the obtained potentials is closely related to the fact that for the most values of the parameters, reflectionless P\"{o}schl-Teller potential has more then one negative mode, i.e.\ the instanton has more then one oscillation. This behavior is known from the context of instantons on the dynamical gravitational background \cite{Hackworth:2004xb}, and usually referred as oscillating instantons.
Physically, these instantons can be interpreted as intermediate states in the vacuum decay process~\cite{Battarra:2012vu}.

Despite the interesting properties of non-analytic potentials, we try to find the solvable Schr\"{o}dinger operators, leading to the analytic reconstructed potentials. We achieved this in Section~\ref{sec:sip}, by applying the reconstruction procedure to the so-called Natanzon operators \cite{Natanzon:1979sr}, which can be thought as the most general Schr\"{o}dinger operators, whose solutions are expressed in terms of the hypergeometric function. We derive the general form of the reconstructed potential and then focus on the shape-invariant operators \cite{Dabrowska:1987fd,Cooper:1986tz,Cooper:1994eh}, specifically Manning-Rosen, hyperbolic P\"{o}schl-Teller and Scarf operators, which constitute the subclass of the Natanzon ones. We fix its parameters, requiring the analyticity of the reconstructed potentials. By doing this way, we found several infinite families of smooth potentials. The latter generalize the known tunneling potentials, but have similar characteristic shapes, namely inverted and non-inverted double-well potentials, cubic-well and sine-Gordon ones. However, some of the have rather different form, namely Manning-Rosen operators allow to obtain triple-well and non-symmetric double-well potentials, whereas Scarf operators leads to the potentials obeying unusual quasi-periodicity property (\ref{eq:nat_S_shift}). Although the obtained potentials have rather simple shapes, the most of them are defined in a parametric form, which can be made explicit only for the specific values of the parameters (cf.~(\ref{eq:nat_MR_expl}), (\ref{eq:nat_hPT_expl}) and (\ref{eq:nat_S_expl})).

It would be interesting to generalize the reconstruction procedure to the case of many (or infinite) degrees of freedom. Particularly, this can be done simply by adding the dependence on spatial coordinates and adding the gradient terms to the action, such that the instanton solution becomes the soliton one in the underlying field theory. Less trivial but more promising generalization is to the case of gauge theories like the Yang-Mills theory, where one has more than one zero mode on the instanton background.

It is important to mention that our work is not the first attempt to generalize the well-known tunneling potentials like double-well one. Among others, it is worth mentioning the so-called Chebyshev wells \cite{Basar:2017hpr,Raman:2020sgw}, which generalize the double-well potentials, from the point of view of (classical and quantum) periods the underlying system. Namely these potentials have the property that these periods obey the second-order differential equation of the hypergeometric type with respect to energy variable, as in the case of double-well, cubic-well, and sine-Gordon potentials. In particular, this means that these periods span the two-dimensional space, and have quite simple but still nontrivial modular properties, similar to the previously known cases, despite the fact that the Chebyshev wells can have an arbitrary number of the degenerate minima.

\section*{Acknowledgments}
The authors express their especial thanks to Nikita Zaigraev for the collaboration at the early stages of this project. The authors are also grateful to Pavel Suprun and Maxim Reva for useful discussions. 
The research was supported by the Russian Science Foundation grant No.\ 23-12-00051, \url{https://rscf.ru/en/project/23-12-00051/}.
	
\appendix
\renewcommand{\thesection}{\Alph{section}}
\renewcommand{\thesubsection}{\Alph{section}.\arabic{subsection}}
\renewcommand{\theequation}{\Alph{section}.\arabic{equation}}

\section{Turning points types} \label{sec:app_tp_types}

	It turns out that the structure of the function $W(\tau)$ defining the fluctuation operator $-\partial_\tau^2 + W(\tau)$ is mostly defined by the structure of turning points reached by the saddle point solution $x_c(\tau)$. For this reason we perform a classification of possible turning points of $x_c(\tau)$.

	\begin{figure}[h!]
	\centering
	\begin{tikzpicture}[scale=1.3]
		\begin{scope}[ultra thick]
			\draw[RoyalBlue] (-0.5,0) -- (2,0);
			\draw[Red] (2,0) -- (2.5,0);
			\draw[densely dotted,Red] (2.5,0) -- (3.5,0);
			\draw[Red] (3.5,0) -- (4,0);
			\draw[ForestGreen] (4,0) -- (8,0);
			\draw[RoyalBlue, -stealth] (8,0) -- (10.5,0){};
			\draw (10.5,0) node[label={[label distance=0]below:$\gamma$}] {};
			\draw (2,0.05) -- (2,-0.05) node[label={[label distance=-4]below:$-2$}] {};
			\draw (4,0.05) -- (4,-0.05) node[label={[label distance=-4]below:$0$}] {};
			\draw[ForestGreen] (6,0.05) -- (6,-0.05);
			\draw (6,-0.05) node[label={[label distance=-4]below:$1$}] {};
			\draw (8,0.05) -- (8,-0.05) node[label={[label distance=-4]below:$2$}] {};
			\draw[thick] (-0.2, 2.2) -- (4.1, 2.2) -- (4.1, 0.8) -- (-0.2, 0.8) -- cycle;
			\draw[RoyalBlue] (0,2) -- (1,2);
			\draw (1,2) node[label={[label distance=0]right:Infinite-time motion}] {};
			\draw[ForestGreen] (0,1.5) -- (1,1.5);
			\draw (1,1.5) node[label={[label distance=0]right:Finite-time motion}] {};
			\draw[Red] (0,1) -- (1,1);
			\draw (1,1) node[label={[label distance=0]right:Forbidden, $\|\eta_0\| = \infty$}] {};
		\end{scope}
		\begin{scope}[thick]
			\draw[-stealth] (0.1,-1) -- (1.9,-1) node[label={[label distance=0]below:$\scriptstyle x$}] {};
			\draw[-stealth] (0.2,-2) -- (0.2,-0.5) node[label={[label distance=0]right:$\scriptstyle V(x)$}] {};
			\draw[domain=0.3:0.36, smooth, variable=\x,WildStrawberry,densely dotted] plot ({\x}, {-0.01/(\x-0.2)^2-1});
			\draw[domain=0.36:1.1, smooth, variable=\x,WildStrawberry,-stealth] plot ({\x}, {-0.01/(\x-0.2)^2-1});
			\draw[domain=1:1.8, smooth, variable=\x,WildStrawberry] plot ({\x}, {-0.01/(\x-0.2)^2-1});
			\draw[-stealth] (4.1,-1) -- (5.9,-1) node[label={[label distance=0]below:$\scriptstyle x$}] {};
			\draw[-stealth] (4.2,-2) -- (4.2,-0.5) node[label={[label distance=0]right:$\scriptstyle V(x)$}] {};
			\draw[domain=4.2:4.7, smooth, variable=\x,WildStrawberry, samples=100] plot ({\x}, {-0.85*(\x-4.2)^0.4-1});
			\draw[domain=4.6:5.2, smooth, variable=\x,WildStrawberry, samples=100,stealth-] plot ({\x}, {-0.85*(\x-4.2)^0.4-1});
			\draw[domain=5.2:5.8, smooth, variable=\x,WildStrawberry, samples=100,densely dotted] plot ({\x}, {-0.85*(\x-4.2)^0.4-1});
			\draw[-stealth] (6.1,-1) -- (7.9,-1) node[label={[label distance=0]below:$\scriptstyle x$}] {};
			\draw[-stealth] (6.2,-2) -- (6.2,-0.5) node[label={[label distance=0]right:$\scriptstyle V(x)$}] {};
			\draw[domain=6.2:6.8, smooth, variable=\x,WildStrawberry, samples=100] plot ({\x}, {-0.5*(\x-6.2)^1.4-1});
			\draw[domain=6.7:7.3, smooth, variable=\x,WildStrawberry, samples=100,stealth-] plot ({\x}, {-0.5*(\x-6.2)^1.4-1});
			\draw[domain=7.3:7.8, smooth, variable=\x,WildStrawberry, samples=100,densely dotted] plot ({\x}, {-0.5*(\x-6.2)^1.4-1});
			\draw[-stealth] (8.1,-1) -- (9.9,-1) node[label={[label distance=0]below:$\scriptstyle x$}] {};
			\draw[-stealth] (8.2,-2) -- (8.2,-0.5) node[label={[label distance=0]right:$\scriptstyle V(x)$}] {};
			\draw[domain=8.2:8.8, smooth, variable=\x,WildStrawberry, samples=100] plot ({\x}, {-0.3*(\x-8.2)^2.5-1});
			\draw[domain=8.7:9.3, smooth, variable=\x,WildStrawberry, samples=100,stealth-] plot ({\x}, {-0.3*(\x-8.2)^2.5-1});
			\draw[domain=9.3:9.8, smooth, variable=\x,WildStrawberry, samples=100,densely dotted] plot ({\x}, {-0.3*(\x-8.2)^2.5-1});
			\draw[-stealth] (0.1,-3.7) -- (1.9,-3.7) node[label={[label distance=0]below:$\scriptstyle \tau$}] {};
			\draw[-stealth] (0.2,-4) -- (0.2,-2.5) node[label={[label distance=0]right:$\scriptstyle W(\tau)$}] {};
			\draw[domain=0.3:0.36, smooth, variable=\t,WildStrawberry,densely dotted] plot ({\t}, {0.01/(\t-0.2)^2-3.7});
			\draw[domain=0.36:1.8, smooth, variable=\t,WildStrawberry,samples=100] plot ({\t}, {0.01/(\t-0.2)^2-3.7});
			\draw[-stealth] (4.1,-3) -- (5.9,-3) node[label={[label distance=0]below:$\scriptstyle \tau$}] {};
			\draw[-stealth] (4.2,-4) -- (4.2,-2.5) node[label={[label distance=0]right:$\scriptstyle W(\tau)$}] {};
			\draw[domain=4.3:4.6, smooth, variable=\t,WildStrawberry,samples=150] plot ({\t}, {-0.01/(\t-4.2)^2-3});
			\draw[domain=4.6:5.8, smooth, variable=\t,WildStrawberry,densely dotted,samples=100] plot ({\t}, {-0.01/(\t-4.2)^2-3});	
			\draw[-stealth] (6.1,-3.7) -- (7.9,-3.7) node[label={[label distance=0]below:$\scriptstyle \tau$}] {};
			\draw[-stealth] (6.2,-4) -- (6.2,-2.5) node[label={[label distance=0]right:$\scriptstyle W(\tau)$}] {};
			\draw[domain=6.3:6.6, smooth, variable=\t,WildStrawberry,samples=150] plot ({\t}, {0.01/(\t-6.2)^2-3.7});
			\draw[domain=6.6:7.8, smooth, variable=\t,WildStrawberry,densely dotted,samples=100] plot ({\t}, {0.01/(\t-6.2)^2-3.7});
			\draw[-stealth] (8.1,-3.7) -- (9.9,-3.7) node[label={[label distance=0]below:$\scriptstyle \tau$}] {};
			\draw[-stealth] (8.2,-4) -- (8.2,-2.5) node[label={[label distance=0]right:$\scriptstyle W(\tau)$}] {};
			\draw[domain=8.3:8.36, smooth, variable=\t,WildStrawberry,densely dotted] plot ({\t}, {0.01/(\t-8.2)^2-3.7});
			\draw[domain=8.36:9.8, smooth, variable=\t,WildStrawberry,samples=100] plot ({\t}, {0.01/(\t-8.2)^2-3.7});			
		\end{scope}
	\end{tikzpicture}
	\captionsetup{justification=centering}
	\caption{Typical shapes of $V(x)$ for various regions of $\gamma$ together with the corresponding $W(\tau)$. Arrows indicate directions to the turning points. Dotted lines show the regions where the behavior of the functions can be modified by higher-order corrections to (\ref{eq:pot_as}).}
	\label{fig:gamma_reg}
\end{figure}
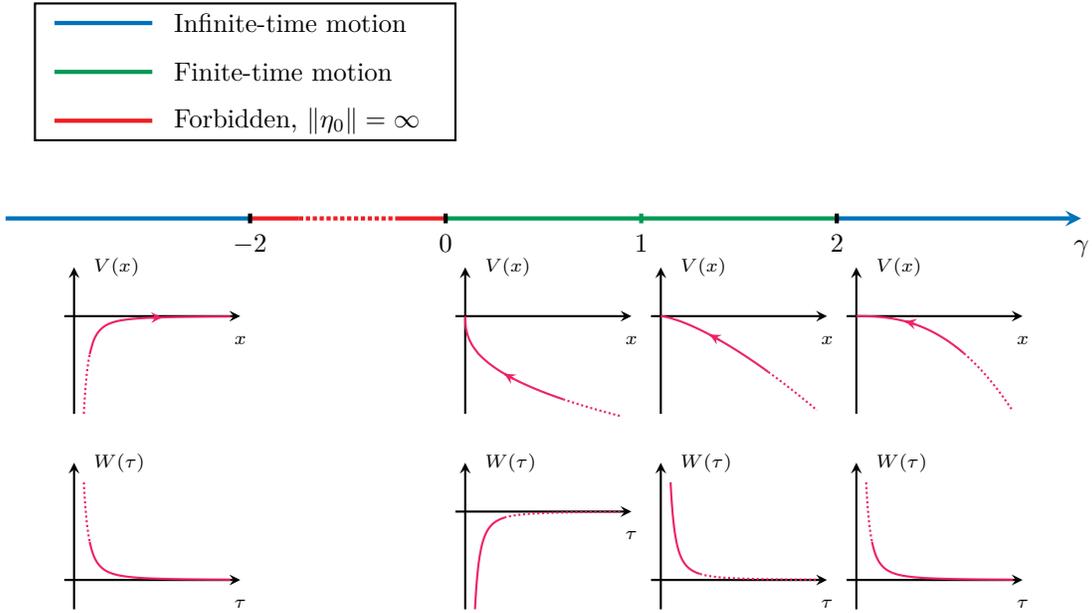

	The first step is to find out a saddle point itself. For this purpose we exploit the energy conservation law (\ref{eq:energy_cons}) and examine it about a near the turning point $x_0$, i.e. a point, such that kinetic term vanishes, i.e.\ $V(x_0) = E$. Without the loss of generality, we set $x_0=0$. Now, let us assume that the potential behaves near the turning point as
	\begin{equation}
		V(x) \sim \frac{\omega^2}2 x^\gamma, \label{eq:pot_as}
	\end{equation}
	so that near this point the energy conservation law reads
	\begin{equation}
		\frac{\dot x^2}2 - \frac{\omega^2}2 x^\gamma = 0, \label{eq:eom_lor_as}
	\end{equation}
	where the constants $\omega$,~$\gamma$ define the leading behavior of the potential near the turning point. Of course, the ``most probable'' case is $\gamma=1$ for which the potential is linear near the turning point. Nevertheless, we will consider the general situation. We also consider negative values of $\gamma$, corresponding to the turning point at infinity $x_0 = \infty$. 
	
	Integration of the equation (\ref{eq:eom_lor_as}) is trivial, so the result, up to unnecessary coefficient, reads
	\begin{equation}
		x_c(\tau) \;\propto\; \tau^{\frac{2}{2-\gamma}}, \label{eq:saddle_as}
	\end{equation}
	where $\gamma \ne 2$. Zero mode $\eta_0(\tau) \propto \dot x_c(\tau)$, hence $W(\tau) = \ddot{\eta}_0(\tau) / \eta_0(\tau)$ behave as
	\begin{equation}
		\eta_0(\tau) \propto \tau^{\frac{\gamma}{2-\gamma}}, \qquad 
		W(\tau) = \frac{2\gamma(\gamma-1)}{(\gamma-2)^2} \frac{1}{\tau^2}.
	\end{equation}
	In the case $\gamma = 2$, the saddle point solution (\ref{eq:saddle_as}) is replaced by
	\begin{equation}
		x_c(\tau) \; \propto \; e^{- \omega \tau},
	\end{equation}
	where the instanton reaches the turning point at $t\to +\infty$ limit. Zero mode and the potential have the form
	\begin{equation}
		\eta_0(\tau) \propto e^{-\omega \tau}, \qquad
		W(\tau) = \omega^2.
	\end{equation}
	
	Let us consider different values of $\gamma$ in order, and select admissible values of $\gamma$. The main criterion is the normalizability of zero mode $\eta_0(\tau)$.
	\begin{itemize}
		\item $\gamma > 0$. In this case the turning point is $x_0=0$. Depending on the sign of the power $2/(2-\gamma)$ of $\tau$ in (\ref{eq:saddle_as}), the turning point is reached in finite or infinite time.
		\begin{itemize}
			\item $\gamma < 2$. In this case the turning point is reached in a finite time at $\tau \to 0$. The normalizability condition of $\eta_0(\tau)$ in the vicinity of $\tau=0$ reads
			\begin{equation}
				\frac{2\gamma}{2-\gamma} > -1,
			\end{equation}
			which is already satisfied due to $0 < \gamma < 2$. It worth noting that $W(\tau)$ change the sign at $\gamma = 1$, so $W(\tau)$ is negative for $0 < \gamma < 1$ and positive for $1 < \gamma < 2$. For $\gamma=1$ the potential $W(\tau)$ vanishes, but higher corrections to (\ref{eq:pot_as}) can modify this property.
			\item $\gamma > 2$. In this case the turning point is reached in an infinite time, i.e.\ at $\tau \to \infty$. The normalizability condition of $\eta_0(\tau)$ in the vicinity of $\tau=\infty$ reads
			\begin{equation}
				\frac{2\gamma}{2-\gamma} < -1,
			\end{equation}
			which is satisfied due to $\gamma > 2$ again. The potential $W(\tau)$ is positive in this case.
		\end{itemize}
		\item $\gamma < 0$. In this case the turning point is at infinity $x_0=\infty$, and it is reached in infinite time $\tau\to \infty$. The normalizability condition of $\eta_0(\tau)$ in the vicinity of $\tau=\infty$ reads
		\begin{equation}
			\frac{2\gamma}{2-\gamma} < -1,
		\end{equation}
		so, one should demand
		\begin{equation}
			\gamma < -2.
		\end{equation}
		\item $\gamma = 2$. In this case the turning point $x_0 = 0$ is reached in the limit $\tau = \infty$. Zero mode decays exponentially in this limit and hence is normalizable.
	\end{itemize}
	The above analysis is summarized at Fig.~\ref{fig:gamma_reg}.

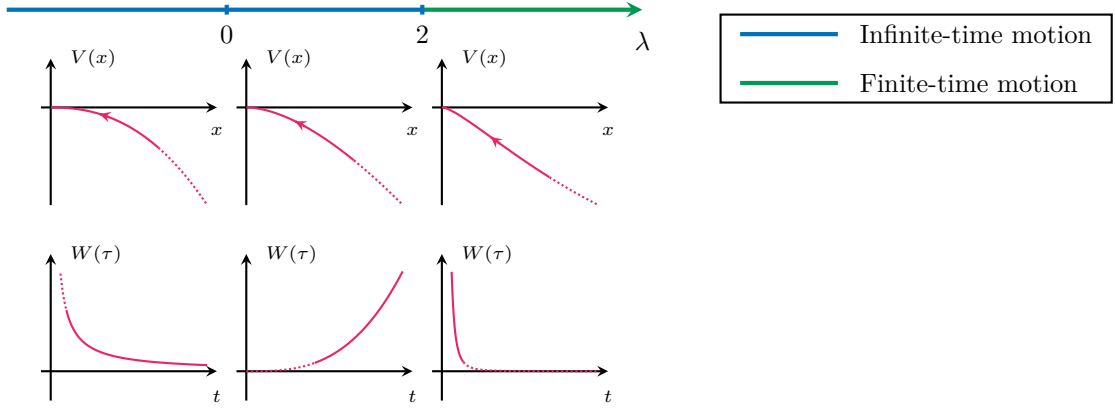
\begin{figure}[h!]
	\centering
	\begin{tikzpicture}[scale=1.3]
		\begin{scope}[ultra thick]
			\draw[RoyalBlue] (-0.25,0) -- (4,0);
			\draw[ForestGreen, -stealth] (4,0) -- (6.25,0){};
			\draw (6.25,0) node[label={[label distance=0]below:$\lambda$}] {};
			\draw[RoyalBlue] (2,0.05) -- (2,-0.05);
			\draw (2,-0.05) node[label={[label distance=-4]below:$0$}] {};
			\draw[RoyalBlue] (4,0.05) -- (4,-0.05);
			\draw (4,-0.05) node[label={[label distance=-4]below:$2$}] {};
			\begin{scope}[xshift=7.25cm,yshift=-1.75cm]
				\draw[thick] (-0.2, 1.7) -- (3.8, 1.7) -- (3.8, 0.8) -- (-0.2, 0.8) -- cycle;
				\draw[RoyalBlue] (0,1.5) -- (1,1.5);
				\draw (1,1.5) node[label={[label distance=0]right:Infinite-time motion}] {};
				\draw[ForestGreen] (0,1) -- (1,1);
				\draw (1,1) node[label={[label distance=0]right:Finite-time motion}] {};
			\end{scope}
		\end{scope}
		\begin{scope}[thick]
			\draw[-stealth] (0.1,-1) -- (1.9,-1) node[label={[label distance=0]below:$\scriptstyle x$}] {};
			\draw[-stealth] (0.2,-2) -- (0.2,-0.5) node[label={[label distance=0]right:$\scriptstyle V(x)$}] {};
			\draw[domain=0.201:0.8, smooth, variable=\x,WildStrawberry, samples=100] plot ({\x}, {-0.988*(\x-0.2)^2/(-ln(\x-0.2)+3)-1});
			\draw[domain=0.7:1.3, smooth, variable=\x,WildStrawberry, samples=100,stealth-] plot ({\x}, {-0.988*(\x-0.2)^2/(-ln(\x-0.2)+3)-1});
			\draw[domain=1.3:1.8, smooth, variable=\x,WildStrawberry, samples=100, densely dotted] plot ({\x}, {-0.988*(\x-0.2)^2/(-ln(\x-0.2)+3)-1});
			\draw[-stealth] (2.1,-1) -- (3.9,-1) node[label={[label distance=0]below:$\scriptstyle x$}] {};
			\draw[-stealth] (2.2,-2) -- (2.2,-0.5) node[label={[label distance=0]right:$\scriptstyle V(x)$}] {};
			\draw[domain=2.201:2.8, smooth, variable=\x,WildStrawberry, samples=100] plot ({\x}, {-0.154*(\x-2.2)^2*(-ln(\x-2.2)+3)-1});
			\draw[domain=2.7:3.3, smooth, variable=\x,WildStrawberry, samples=100,stealth-] plot ({\x}, {-0.154*(\x-2.2)^2*(-ln(\x-2.2)+3)-1});
			\draw[domain=3.3:3.8, smooth, variable=\x,WildStrawberry, samples=100, densely dotted] plot ({\x}, {-0.154*(\x-2.2)^2*(-ln(\x-2.2)+3)-1});
			\draw[-stealth] (4.1,-1) -- (5.9,-1) node[label={[label distance=0]below:$\scriptstyle x$}] {};
			\draw[-stealth] (4.2,-2) -- (4.2,-0.5) node[label={[label distance=0]right:$\scriptstyle V(x)$}] {};
			\draw[domain=4.201:4.8, smooth, variable=\x,WildStrawberry, samples=100] plot ({\x}, {-0.0241*(\x-4.2)^2*(-ln(\x-4.2)+3)^3-1});
			\draw[domain=4.7:5.3, smooth, variable=\x,WildStrawberry, samples=100,stealth-] plot ({\x}, {-0.0241*(\x-4.2)^2*(-ln(\x-4.2)+3)^3-1});
			\draw[domain=5.3:5.8, smooth, variable=\x,WildStrawberry, samples=100, densely dotted] plot ({\x}, {-0.0241*(\x-4.2)^2*(-ln(\x-4.2)+3)^3-1});
			\draw[-stealth] (0.1,-3.7) -- (1.9,-3.7) node[label={[label distance=0]below:$\scriptstyle t$}] {};
			\draw[-stealth] (0.2,-4) -- (0.2,-2.5) node[label={[label distance=0]right:$\scriptstyle W(\tau)$}] {};
			\draw[domain=0.3:0.36, smooth, variable=\t,WildStrawberry,densely dotted] plot ({\t}, {0.1/(\t-0.2)-3.7});
			\draw[domain=0.36:1.8, smooth, variable=\t,WildStrawberry,samples=100] plot ({\t}, {0.1/(\t-0.2)-3.7});
			\draw[-stealth] (2.1,-3.7) -- (3.9,-3.7) node[label={[label distance=0]below:$\scriptstyle t$}] {};
			\draw[-stealth] (2.2,-4) -- (2.2,-2.5) node[label={[label distance=0]right:$\scriptstyle W(\tau)$}] {};
			\draw[domain=2.2:2.9, smooth, variable=\t,WildStrawberry,samples=150,densely dotted] plot ({\t}, {0.25*(\t-2.2)^3-3.7});
			\draw[domain=2.9:3.8, smooth, variable=\t,WildStrawberry,samples=100] plot ({\t}, {0.25*(\t-2.2)^3-3.7});
			\draw[-stealth] (4.1,-3.7) -- (5.9,-3.7) node[label={[label distance=0]below:$\scriptstyle t$}] {};
			\draw[-stealth] (4.2,-4) -- (4.2,-2.5) node[label={[label distance=0]right:$\scriptstyle W(\tau)$}] {};
			\draw[domain=4.3:4.42, smooth, variable=\t,WildStrawberry,samples=100] plot ({\t}, {0.001/(\t-4.2)^3-3.7});
			\draw[domain=4.42:5.8, smooth, variable=\t,WildStrawberry,samples=100,densely dotted] plot ({\t}, {0.001/(\t-4.2)^3-3.7});				
		\end{scope}
	\end{tikzpicture}
	\captionsetup{justification=centering}
	\caption{Typical shapes of $V(x)$ for various regions of $\lambda$ together with the corresponding $W(\tau)$. Arrows indicate directions to the turning points. Dotted lines show the regions where the behavior of the functions can be modified by higher-order corrections to (\ref{eq:pot_log_as}).}
	\label{fig:lambda_reg}
\end{figure}
	
	The presence of the values of $\gamma$, where the behavior of the saddle point solution and the corresponding $W(\tau)$ is changed discontinuously, e.g. $\gamma=2$, motivate as to enlarge the class of the potentials (\ref{eq:pot_as}) near the turning point to obtain smoother transition from different regimes. For this purpose we generalize (\ref{eq:pot_as}) by the logarithmic correction
	\begin{equation}
		V(x) \sim  \frac{\omega^2}2 x^\gamma (-\log x)^\lambda, \qquad x \to 0. \label{eq:pot_log_as}
	\end{equation}
	
	Now, let us consider the case $\gamma = 2$. As before, the solution can be extracted from the energy conservation law
	\begin{equation}
		\frac{\dot x^2}2 - \frac{\omega^2}2 x^\gamma (-\log x)^\lambda = 0,
	\end{equation}
	and reads
	\begin{equation}
		x_c(\tau) \propto \exp \left[ - (\pm \tilde \omega \tau)^{\frac{2}{2-\lambda}} \right], \qquad \tilde \omega = \frac{2}{|2 - \lambda|} \omega,
	\end{equation}
	for $\lambda \ne 2$. Zero mode reads
	\begin{equation}
		\eta_0(\tau) \propto (\pm \tilde \omega \tau)^{\frac{\lambda}{2-\lambda}} \exp \left[ - (\pm \tilde \omega \tau)^{\frac{2}{2-\lambda}} \right],
	\end{equation}
	whereas the potential $W(\tau)$ has the following form
	\begin{equation}
		W(\tau) = \tilde\omega^2 \left[ (\alpha-1)(\alpha-2) (\tilde\omega \tau)^{-2} - 3(\alpha-1) (\tilde\omega \tau)^{\alpha-2} + \alpha^2 (\tilde\omega \tau)^{2\alpha - 2} \right], \qquad \alpha = \frac{2}{2-\lambda}, \label{eq:W_pot_g_2}
	\end{equation}
	where we keep only the plus sign. As we will see, not all terms remain important, in $x(\tau) \to 0$ limit.
	
	In the case $\lambda = 2$ the turning point solution is as follows
	\begin{equation}
		x_c(\tau) = e^{- e^{\pm \omega t}}
	\end{equation}
	Zero mode and $W(\tau)$ have the following form
	\begin{equation}
		\eta_0(\tau) \propto e^{\pm \omega t} e^{- e^{\pm \omega t}}, \qquad
		W(\tau) = \omega^2 \left[(e^{\pm \omega t} - \tfrac32)^2 - \tfrac54\right].
	\end{equation}
	
	Now, let us examine the asymptotic behavior of $W(\tau)$, when $x_c(\tau)$ reaches the turning point, for different values of $\lambda$.
	\begin{itemize}
		\item $\lambda < 2$. In this case $x_c(\tau)$ approaches the turning point in an infinite time in $\tau \to \infty$ limit. Thus, the leading term in $W(\tau)$ is those, having the most positive power of $\tau$. This is the latter one, since $\alpha > 0$ for $\lambda < 2$. Hence
		\begin{equation}
			W(\tau) \sim \tilde \omega^2 \alpha^2 (\tilde\omega \tau)^{2\alpha-2}, \qquad \tau \to \infty.
		\end{equation}
		The power of the leading term can be either positive of negative, depending on the value of $\gamma$. More specifically
		\begin{align}
			-2 < 2&\alpha - 2 < 0,  & -\infty < \lambda & {} < 0, \\
			0 < 2&\alpha - 2 < +\infty,  & 0 < \lambda & {} < 2.
		\end{align}
		The case of $\lambda = 0$ was examined before.
		\item $\lambda = 2$. In this case $x_c(\tau)$ approaches the turning point in an infinite time in $\tau \to \infty$ limit too.
		\item $\lambda > 2$. In this case $x_c(\tau)$ approaches the turning point in a finite time in $\tau \to 0$ limit. Thus, the leading term in $W(\tau)$ is those, having the most negative power of $\tau$. This is the latter one, since $\alpha < 0$ for $\lambda > 2$. Hence
		\begin{equation}
			W(\tau) \sim \tilde \omega^2 \alpha^2 (\tilde\omega \tau)^{2\alpha-2}, \qquad \tau \to 0.
		\end{equation}
		The range of the power of $\tau$ in $W(\tau)$ reads
		\begin{equation}
			-\infty < 2\alpha-2 < -2.
		\end{equation}
	\end{itemize}
	These results is summarized at Fig.~\ref{fig:lambda_reg}.
	
	The above analysis allows to make general conclusions about the correspondence between the fluctuation operator, specifically function $W(\tau)$, defining it, on the one hand side, and the reconstructed potential on the other side. 
	
	Specifically, if near the turning point the potential has the leading behavior $V(x) \propto x^\gamma$, the corresponding function $W(\tau)$ can have $1/\tau^2$-singularity, $\tau\to 0$ if $0<\gamma<1$ and $1<\gamma<2$. Since we discuss asymptotic behavior, this particularly means the converse, namely if $W(\tau)$ has $1/\tau^2$-singularity for $\tau \to 0$, then the reconstructed potential has the leading behavior $V(x) \propto x^\gamma$ with $0<\gamma<2$, $\gamma\ne 1$ near the turning point, i.e.\ at leading order, the potential is non-analytic as the function of the complexified variable $x$. In these discussion we exclude ``fall to the center'' situation (see e.g.\ \cite{perelomov1970fall}), in which the prefactor of $1/\tau^2$, in the leading behavior $W(\tau)$ is less then $-1/4$, corresponding to complex values of $\gamma$. 
	
	Moreover, if the leading behavior of the potential has the logarithmic correction to $\gamma=2$, i.e. $V(x) \propto x^2 (-\log x)^\lambda$, then $W(\tau)$ has the power-law asymptotics $W(\tau)\propto \tau^{2\lambda/(\lambda-2)}$. Conversely, this means that, if $W(\tau)$ has the power-law asymptotics $W(\tau) \propto \tau^\beta$, $\beta>0$, then the leading-order behavior of the reconstructed potential has the logarithmic singularity $V(x) \propto x^2 (-\log x)^{2\beta/(\beta+2)}$ near the turning point.

\section{Analytic structure of reconstructed potential}\label{sec:app_analytic}

\subsection{General case} \label{sec:app_tp_general}
	Let us discuss the local behavior of the potentials $V(x)$, defined parametrically in (\ref{eq:pot_rec_par}), namely its analytic behavior as its argument $x$ (rather than the parameter $\tau$). Suppose zero mode $\eta_0(\tau)$ has Taylor expansion near some point $\tau_0$
	\begin{equation} \label{eq:app_eta_sr}
		\eta_0(\tau) = \sum_{k=0}^\infty p_k \zeta^k, \qquad \zeta\coloneqq \tau-\tau_0.
	\end{equation}
	Then, the instanton trajectory has the following Taylor expansion
	\begin{equation} \label{eq:app_x_sr}
		x_c(\tau) = \int^\tau d\tau \, \eta_0(\tau) = x_0 + \sum_{k=0}^\infty \frac{p_k}{k+1} \zeta^{k+1}, \qquad
		x_0\coloneqq x_c(\tau_0),
	\end{equation}
	where for simplicity we set $\nu = 1$. If instanton trajectory has nonvanishing velocity $\dot x_c(\tau_0)= \eta_0(\tau_0) \ne0$, which implies $p_0\ne 0$, then the inverse series for $\zeta$ in integer powers of $x-x_0$ exist, and the substitution to series expansion (\ref{eq:app_eta_sr}) for zero mode $\eta_0(\tau)$ results into well-defined Taylor series in integer powers $x-x_0$ for the potential $V(x_c(\tau)) = 1/2 \, (\eta_0(\tau))^2$. Thus, if at some the instanton trajectory point is smooth and has nonzero velocity, the potential is also smooth function of coordinate near the latter point.
	
	The situation is different, if the instanton trajectory has vanishing velocity $\dot x_c(\tau_0)$, i.e.\ exhibits a turning point $x_{\mathrm{tp}} = x_c(\tau_0)$. Vanishing velocity implies $p_0 = 0$, and assuming that the root is simple, i.e.\ $p_1 \ne 0$ we conclude that the inverse series for $\zeta$ is in half-integer powers of $x - x_{\mathrm{tp}}$
	\begin{equation}
		\zeta = \sum_{k=1}^\infty q_k \, (x-x_{\mathrm{tp}})^{k/2},
	\end{equation}
	where, for instance
	\begin{equation}
		q_1 = \frac{\sqrt{2}}{\sqrt{p_1}}, \qquad
		q_2 = -\frac{2 p_2}{3 p_1^2}, \qquad
		q_3 = \frac{10 p_2^2-9 p_1 p_3}{9 \sqrt{2}p_1^{7/2}}.
	\end{equation}
	Subsequent substitution to the expression for zero mode and to the parametric form of the potential, we obtain
	\begin{equation} \label{eq:app_pot_sing}
		V(x) = p_1 (x-x_{\mathrm{tp}}) \left(1 + \frac{4\sqrt{2}}{3} \frac{p_2}{p_1} \sqrt{\frac1{p_1}(x-x_{\mathrm{tp}})} + \frac{9p_1p_3 - 2 p_2^2}{3 p_1^2} \frac1{p_1} (x-x_{\mathrm{tp}}) + \ldots   \right),
	\end{equation}
	where ellipses contain the remaining part of the series in half-integer power of $x-x_{\mathrm{tp}}$. Thus, if instanton trajectory has an intermediate turning point, and its velocity as the function of $\tau$ has simple root at this point, then the potential generally has a square root singularity at this point. 
	
	The situation can be saved if the zero mode, i.e.\ velocity of instanton trajectory is an odd function of $\tau$ about the turning point. This means that $p_{2k} = 0$ in (\ref{eq:app_eta_sr}), hence the instanton trajectory (\ref{eq:app_x_sr}) has series in even powers of $\zeta$. In this case the equality can be solved for $\zeta^2$ as
	\begin{equation} \label{eq:app_z_pw}
		\zeta^2 = \sum_{k=1}^\infty r_k \, (x-x_{\mathrm{tp}})^k,
	\end{equation}
	where first few coefficients read
	\begin{equation}
		r_1 = \frac{2}{p_1}, \qquad
		r_2 = -\frac{2 p_3}{p_1^3}, \qquad
		r_3 = \frac{12 p_3^2-8 p_1 p_5}{3 p_1^5}.
	\end{equation}
	Once $p_{2k}=0$ in (\ref{eq:app_eta_sr}), the potential $V(x_c(\tau)) = 1/2 \, (\eta_0(\tau))^2$ is the function of even powers of $\zeta$. Thus, the substitution of (\ref{eq:app_z_pw}) gives the series in integer powers of $x-x_{\mathrm{tp}}$
	\begin{equation}
		V(x) = p_1 (x-x_{\mathrm{tp}}) \left( 1+\frac{3 p_3 }{p_1^2}(x-x_{\mathrm{tp}})+\frac{2 (10 p_1 p_5-3 p_3^2)}{3 p_1^4}(x-x_{\mathrm{tp}})^2 + \ldots \right),
	\end{equation}
	hence we conclude that $V(x)$ is an analytic function near the underlying turning point.
	
	Thus, we can make a conclusion on the analytic properties of the reconstructed potential strictly inside the range covered by the instanton trajectory. The discussion of the boundary points, corresponding to $\tau\to\pm\infty$ is more subtle and will be done under additional assumptions below. 
	
	Given the specific fluctuation operator together with its bound state eigenfunctions and spectrum, we can always shift it by a constant such that its $n$-th bound state eigenfunction has zero eigenvalue, i.e.\ becomes a zero mode. According to oscillation theorem, $n$-th excite bound state eigenfunction has exactly $n$ internal roots. As a result, using the above analysis, we conclude that if the zero mode corresponds to the ground state, then the reconstructed potential is necessarily analytic within the instanton trajectory. Further, if the zero mode corresponds to the first excited state, then the potential is analytic only if the zero mode is odd about its zero. This zero exactly corresponds to the intermediate turning point of the instanton trajectory. Higher excited states have no chance of leading to analytic reconstructed potential, because the function having more then one, but a finite number of zeros cannot be an odd function about all these zeros. Nevertheless, even in this case the instanton trajectory still can have one turning point, where the potential is analytic, besides other nonanalytic ones.
	
\subsection{Analytic structure of $V_{\ell,m}$} \label{sec:app_tp_PT}
	Here we examine the analytic properties of the potentials $V_{\ell,m}(x)$, parametrically defined in (\ref{eq:PT_pot_par}) near the turning points.
	
	Let us first discuss the analytic structure inside the range of instanton trajectory, applying the results of the previous subsection. According to (\ref{eq:pt_zero_mode}), zero mode is defined via the associated Legendre polynomial as $\eta_{0}(\tau) = C_0 P_\ell^m(\tanh \tau)$. Thus, we are interested in the zeros of $P_\ell^m(y)$, lying inside the range $-1 < y <1$. The associated Legendre polynomial $P_\ell^m(y)$ has exactly $\ell-m$ simple zeros, and its parity, as the function of $y$, coincides with the parity of $\ell-m$. Thus, using the above analysis, we conclude that the potential is analytic inside the range of instanton trajectory in two cases, namely if $m = \ell$, i.e.\ the zero mode coincides with the ground state eigenfunction, and if $m = \ell - 1$, i.e.\ zero mode is first exited state and the odd function of $\tau$. For $\ell-m > 1$ the potential necessarily has square root singularities (\ref{eq:app_pot_sing}). Nevertheless even in this case, for odd $\ell-m$ there is the intermediate turning point, corresponding to $y=0$ at which the potential is analytic.

	To examine the analytic behavior of the potential near the boundary turning points, corresponding to $y=\pm1$ in (\ref{eq:PT_pot_par}), we need extend the analysis of the previous subsection. The main technical difference is that near these points associated Legendre polynomials are not linear at the leading order, but have the following power-law dependence
	\begin{equation} \label{eq:pt_p_of_xi}
		P(y) = \nu C_0 P_\ell^m(y) = \xi^{m/2} \sum_{k=0}^{\infty} q_k \, \xi^k, \qquad \xi = \frac12(1 - y), 
	\end{equation}
	so that integration in (\ref{eq:pt_traj}) gives
	\begin{equation} \label{eq:pt_x_of_xi}
		x -x_{\mathrm{tp}} = \xi^{m/2} \sum_{k=0}^\infty \tilde{q}_k \, \xi^k, \qquad \tilde{q}_k = -\sum_{j=0}^k\frac{q_k}{m+2k}.
	\end{equation}	
	This equality can be inverted in the from of formal power series
	\begin{equation} \label{eq:pt_xi_of_x}
		\xi = \sum_{k=1}^{\infty} \chi_{k} \, (x-x_{\mathrm{tp}})^{\frac{2k}m},
	\end{equation}
	where without loss of generality we assume that $\tilde{q}_0=-q_0/m$ is positive, so at leading order $x-x_{\mathrm{tp}}$ is positive too. The coefficients $\chi_k$ can be found order by order by substituting (\ref{eq:pt_x_of_xi}) into (\ref{eq:pt_xi_of_x}). The first two coefficients read
	\begin{equation}
		\chi_1 = \left(-\frac{m}{q_0}\right)^{\!\frac{2}{m}},\qquad
		\chi_2 = -\frac{2}{m+2}\left(-\frac{m}{q_0}\right)^{\!\frac{4}{m}}\frac{q_0+q_1}{q_0}.
	\end{equation}
	Substitution to (\ref{eq:pt_p_of_xi}), and then to (\ref{eq:PT_pot_par}) gives the expression for the potential
	\begin{equation}
		V_{\ell,m}(x) = \frac{m^2}2 (x-x_{\mathrm{tp}})^2 \left(1+ \frac{2(2q_1-m q_0)}{(m+2)q_0}\Bigl[-\frac{m}{q_0}(x-x_{\mathrm{tp}})\Bigr]^{\frac{2}m} + \ldots \right),
	\end{equation}
	where ellipses denote the remaining series in powers of $(x-x_{\mathrm{tp}})^{2/m}$. Fixing the normalization coefficient $\nu$ such that $\tilde{q}_0=-q_0/m=1$ and using the explicit form of the associated Legendre polynomial in (\ref{eq:pt_p_of_xi}), we obtain a more explicit form (\ref{eq:pt_pot_bdy_tp}).
	Thus, we see that the potential has asymptotically harmonic behavior near the boundary turning point. At the same time, for $m>2$ the subleading corrections have $2/m$-th root behavior, hence nonanalytic at these points.
\subsection{Analytic structure of $V_N$}
\label{sec:app_tp_nat}
	In this Appendix, we examine the analytic behavior of the potential $V_N(x)$, defined parametrically in (\ref{eq:nat_sip_pot}).
	As discussed in Appendix~\ref{sec:app_tp_general}, such defined potential can exhibit nonanalyticity only at the boundary turning points of the instanton trajectory, since the underlying zero mode is chosen to be the ground state of the fluctuation operator. Thus, we will invert the function (\ref{eq:nat_inst_traj}) defining the instanton trajectory $x(z)$ in terms of suitable power series, substitute to the parametric expression for the potential, and examine which range of the parameters $r_0$,~$r_1$ leads to the analytic behavior near the turning points.
	
	The turning point can only correspond to the values $z=0$, $z=1$, and $z=\infty$ of the instanton trajectory $x(z)$. The latter can be expressed in terms of hypergeometric function, and have three different forms (\ref{eq:nat_inst_traj_0})--(\ref{eq:nat_inst_traj_infty}) near $z=0$, $z=1$ and $z=\infty$, respectively. Thus, we can use the appropriate expansion of the hypergeometric function near these point to invert the function $x(z)$ in terms of power series. Near $z = 0$ suitable power series have the form
	\begin{equation}\label{eq:app_tp_nat_z_sr_0}
		z(x) = (\delta X_0)^{\frac1{r_0}}\sum_{k=0}^\infty s_k \, (\delta X_0)^{\frac{k}{r_0}},
	\end{equation}
	where $\delta X_0 = \tilde{\nu}^{-1} r_0 \, (x - x_0)  $ defines the distance from the corresponding turning point $x_0$. Substituting to (\ref{eq:nat_inst_traj_0}), we find first few coefficients of the series
	\begin{equation}
		s_0 = 1, \qquad
		s_1 = -\frac{1-r_0}{1+r_0}, \qquad
		s_2 = \frac{\left(r_1-1\right) \left(r_0^2+\left(3 r_1-1\right) r_0+5 r_1-4\right)}{2 \left(r_0+1\right){}^2 \left(r_0+2\right)},
	\end{equation}
	so the potential near the turning point have the form
	\begin{align}
		V_N(x) = \frac{2 \tilde{\nu}^2}{A} (\delta X_0)^{2-\frac{q_0}{r_0}} \left(1 - \left[(2r_0-q_0)\frac{1-r_1}{1+r_0} + (2r_1-q_1)\right] (\delta X_0)^{\frac1{r_0}} + \ldots \right)
	\end{align}
	where ellipses contain the remaining part of the series in powers of $(\delta X_0)^{1/{r_0}}$. Thus, we conclude that for the potential to be analytic near the turning point corresponding to $x_0$, $1/r_0$ should be a positive integer, say $r_0 = 1/N_0$.
	Very similar analysis, applied to (\ref{eq:nat_inst_traj_1}) shows that $V_N(x)$ is analytic near the turning point, corresponding to $z=1$, one should require that $r_1=1/N_1$, where $N_1$ is a positive integer.
	
	To analyze the turning point corresponding to $z=\infty$, we should invert (\ref{eq:nat_inst_traj_infty}). This can be done using the series expansion, slightly different from (\ref{eq:app_tp_nat_z_sr_0}), namely
	\begin{equation}
		\frac1{z(x)} = (\delta X_\infty)^{\frac1{1-r_0-r_1}}\sum_{k=0}^\infty t_k \, (\delta X_\infty)^{\frac{k}{1-r_0-r_1}},
	\end{equation}
	where $\delta X_\infty = -\tilde{\nu}^{-1} e^{\mp i \pi r_0}(1-r_0-r_1)  \, (x - x_\infty)$. As in the previous case, substituting to (\ref{eq:nat_inst_traj_infty}), we can find the values of the series coefficients. Subsequent substitution to the expression for the potential has the form of the expansion in the powers of $(\delta X_\infty)^{1/{1-r_0-r_1}}$, whereas the overall coefficient equals to $\delta X_\infty^{2-{(2-q_0-q_1)}/{(1-r_0-r_1)}}$ up to a constant. Thus the analyticity of the potential near the turning point, corresponding to $z=\infty$, requires $1-r_0-r_1 = 1/M$ where $M$ is a positive integer.

\bibliography{main}

\end{document}